\documentclass[
a4paper
,prd
,nofootinbib
,preprint
,tightenlines
,superscriptaddress
]{revtex4}
\usepackage{makecell}
\usepackage{braket}
\usepackage{amsmath}
\usepackage{eufrak}
\usepackage{graphicx}
\usepackage{subfigure}
\usepackage{grffile}
\usepackage{float}
\usepackage[a4paper, total={6.5in, 9in}]{geometry}
\usepackage[colorlinks,
linkcolor=blue,
anchorcolor=blue,
citecolor=blue
]{hyperref}

\usepackage{comment}

\newcommand{\eq}[1]{Eq.~\eqref{#1}}
\newcommand{\Sec}[1]{Sec.~\ref{#1}}  \newcommand{\fig}[1]{Fig.~\ref{#1}}
\newcommand{\tab}[1]{Tab.~\ref{#1}}
\newcommand{\App}[1]{Appendix~\ref{#1}}

\newcommand{\REF}[1]{Ref.~\cite{#1}}

\newcommand{\lra}[1]{\left(#1\right)}

\newcommand{\lrd}[1]{\left|#1\right|}
\newcommand{\lre}[1]{\left\langle#1\right\rangle}

\xdef\figsizeThree{5.1cm}
\xdef\figsizeTwo{7.8cm}
\xdef\figsizeOne{8.5cm}
\xdef\Gran{LNGS}
\xdef\Jinp{CJPL}
\xdef\plotsdir{./}

\date{\small\itshape Last update: \today}
\begin{document}
	\title{Diurnal modulation of electron recoils from DM-nucleon scattering through the Migdal effect}
    \author{Mai Qiao}
    \affiliation{CAS key laboratory of theoretical Physics, Institute of
    Theoretical Physics, Chinese Academy of Sciences, Beijing 100190,
    China; \\ School of Physics, University of Chinese Academy of
    Sciences, Beijing 100049, China.}
\author{Chen Xia}
\affiliation{Tsung-Dao Lee Institute \& School of Physics and Astronomy, Shanghai Jiao Tong University, China}
    \affiliation{Key Laboratory for Particle Astrophysics and Cosmology (MOE) \& Shanghai Key Laboratory for Particle Physics and Cosmology, Shanghai Jiao Tong University, Shanghai 200240, China}
    \author{Yu-Feng Zhou}
    \affiliation{CAS key laboratory of theoretical Physics, Institute of
    Theoretical Physics, Chinese Academy of Sciences, Beijing 100190,
    China; \\ School of Physics, University of Chinese Academy of
    Sciences, Beijing 100049, China.}
    \affiliation{School of Fundamental Physics and Mathematical Sciences, Hangzhou
    Institute for Advanced Study, UCAS, Hangzhou 310024, China. }
    \affiliation{International Centre for Theoretical Physics Asia-Pacific, Beijing/Hangzhou,
    China.}
  \begin{abstract}
    Halo dark matter (DM) particles could lose energy due to the scattering off nuclei within the Earth before reaching the underground detectors of DM  direct detection experiments. 
This Earth shielding effect can result in diurnal modulation of the DM-induced recoil event rates observed underground due to the self-rotation of the Earth.
For electron recoil signals from DM-electron scatterings, the current experimental constraints are very stringent such that the diurnal modulation cannot be observed for halo DM.
We propose a novel type of diurnal modulation effect: diurnal modulation in {\it electron} recoil signals induced by DM-{\it nucleon} scattering via the Migdal effect. 
We set so far the most stringent constraints on  DM-nucleon scattering cross section via the Migdal effect for sub-GeV DM using the S2-only data of  PandaX-II and PandaX-4T with improved simulations of the Earth shielding effect.
Based on the updated constraints, we show that the Migdal effect induced diurnal modulation of electron events can still be significant in the low energy region, and can be probed by experiments such as PandaX-4T in the near future.
\end{abstract}
\maketitle

\section{Introduction}\label{section_introduction}
Although enormous astrophysical observations have supported the existence of dark matter~(DM) as the dominant form of matter in the present Universe, whether or not DM has non-gravitational interactions with the standard model~(SM) particles is still unknown.
In recent years, many underground DM direct detection~(DD) experiments have been constructed to detect possible signals from halo DM particles scattering off target nuclei or electrons within the detectors.

If DM particles couple to SM particles, the halo DM particles from the space can scatter off the nuclei or electrons within the  Earth before reaching the underground detectors, which leads to direction changes and energy losses of the DM particles, and eventually the deformation of the underground DM energy spectrum.  This effect is usually referred to as the Earth shielding effect, which can be significant if the DM couplings are large enough.
For an underground DD experiment with a given location and depth, as the Earth self-rotates during a sidereal day and orbits around the Sun, the shielding effect varies nearly periodically in a sidereal day and also varies slowly during a year. Consequently, the recoil event rates to be observed by the detector should also vary in the same manner,  which is known as the diurnal modulation effect~\cite{Collar:1992qc, Collar:1993ss, LUX:2018xvj, DAMA-LIBRA:2014lld, Foot:2014osa,Chen:2021ifo}. 
The diurnal modulation, if observed, can provide additional information on  the DM property, and can be helpful in distinguishing DM signals from other time-independent backgrounds.

\begin{figure}[t]
    \centering
    \includegraphics[width=\figsizeOne]{\plotsdir/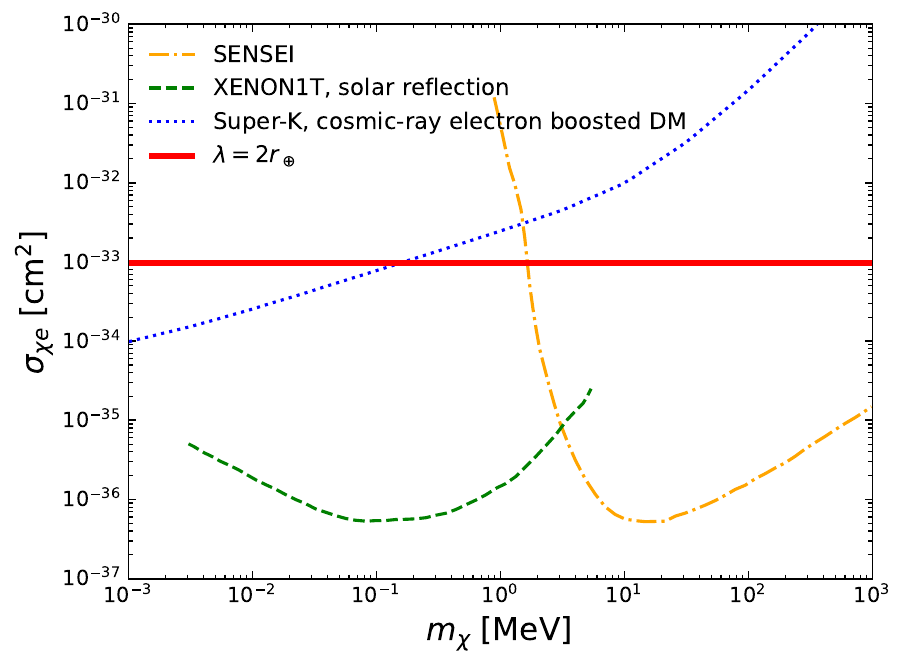}
\caption{
        A selection of current constraints on DM-electron scattering cross sections $\sigma_{\chi e}$  from the data of SENSEI for halo DM~(orange dash-dotted)~\cite{Barak:2020fql}, data of XENON1T on the solar reflection of halo DM~(green dashed)~\cite{An:2017ojc}, and the data of Super-K on cosmic-ray electron boosted DM~(blue dotted)~\cite{Xia:2022tid}.
The DM-electron scattering cross section corresponding to the mean-free-path $\lambda=2r_\oplus$~(red solid) is also shown. 
}
    \label{fig_mean_free_path}
\end{figure}

The diurnal modulation of nuclear recoil event rate for halo DM-nucleus scattering has been investigated extensively (see, e.g. \cite{Collar:1992qc,Collar:1993ss,Bernabei:2015nia}).
The required large DM-nucleon cross sections are, however,  already excluded by the null results of the current DD experiments for DM particle mass above GeV scale~\cite{DarkSide-50:2022qzh,LZ:2022ufs,PandaX-4T:2021bab,XENON:2023sxq}.
For the case of halo DM-electron scatterings, in order to observe the sizable diurnal modulation, the DM-electron scattering cross section $\sigma_{\chi e}$ also needs to be large enough. A rather conservative criterion for the Earth shielding to be effective for DM-electron scatterings is that  the mean-free-path of the DM particle $\lambda=1/{n_e\sigma_{\chi e}}$ (where $n_e$ is the electron number density of the Earth) should be smaller than the diameter of the Earth, 
i.e. $\lambda \lesssim 2 r_\oplus$, where $r_\oplus=6371~{\text{km}}$ is the average radius of the Earth. 
Otherwise, the Earth can be considered to be transparent to the halo DM.
The required DM-electron scattering cross section  can be estimated as
$\sigma_{\chi e} \gtrsim (2 \rho_\oplus r_\oplus \sum_N  f_N Z_N / m_N )^{-1}$,
where $Z_N$ and $m_N$ are the atomic number and mass of a nucleus $N$, respectively, and $f_N$ is the relative abundance of the nucleus $N$ within the Earth. 
For a simple estimation, let us consider a homogeneous Earth model with an average mass density $\rho_\oplus=2.7~{\text{g}/\text{cm}^3}$~\cite{Rudnick:2003xyz}, and the chemical composition of the Earth  dominated by the element $^{16}\text{O}$. For such a simple  Earth model, we find 
$\sigma_{\chi e} \gtrsim \mathcal{O}(10^{-33})~\text{cm}^2$.
Unfortunately, such a large cross section is completely excluded by the current experiments for the DM masses in a very large range. The constraints arise from the analysis on the data of SENSEI for MeV scale halo DM~\cite{Barak:2020fql}, 
on the data of XENON1T for the solar reflected multi-keV scale halo DM~\cite{An:2017ojc}, 
and on the data of Super-K for cosmic-ray electron boosted DM with mass at keV scale and below~\cite{Xia:2022tid}. The current constraints are summarized in \fig{fig_mean_free_path}. 
It can be seen that the diurnal modulation of electron recoil signals from halo DM-electron scattering is unlikely to be significant for DM particles with mass above keV scale.

In this work, we discuss a novel type of diurnal modulation effect: diurnal modulation in {\it electron} recoil signals induced by DM-{\it nucleus} scattering via the Migdal effect. 
The Migdal effect refers to the process in which the recoil of the nucleus inside the atom leads to electron signals due to the ionization or excitation of the atom, which has been considered to lower the detection threshold for DM-nucleus scatterings~\cite{Migdal, Dolan:2017xbu, Ibe:2017yqa, Essig:2019xkx, Bernabei, Aprile:2019xxb,Li:2022acp,Flambaum:2020xxo,Cox:2022ekg}. 
We show that this diurnal modulation effect in electron recoil signals which is an important feature for the Migdal effect can be significant and testable in the ongoing and future experiments.
We first set so far the most stringent constraints on  DM-nucleon scattering cross section via the Migdal effect using the latest S2-only data of PandaX-II and PandaX-4T together with that of XENON10 and XENON1T with improved Monte-Carlo simulations of the Earth shielding effect.
Then, based on the updated constraints, we predict that the Migdal effect induced diurnal modulation of electron recoils can still be significant, and can be observed in experiments such as PandaX-4T in the low energy region in the near future.

This work is organized as follows.
In \Sec{section_migdal_effect}, we briefly review the kinematics and formalism of the Migdal effect of isolated atoms.
In \Sec{section_diurnal_effect}, we calculate the Earth shielding effect of halo DM due to DM-nucleus scatterings and calculate the diurnal modulation of the Migdal effect using the underground DM flux.
In \Sec{section_constraints}, we set the constraints on spin-independent (SI) DM-nucleon scattering cross section using the S2-only data of PandaX-II, PandaX-4T, XENON10, and XENON1T via the Migdal effect with the Earth shielding effect included.
In \Sec{section_projection}, we give predictions for the diurnal modulation in S2-only data in the low energy regions for the PandaX-4T experiment.  
Our main results are summarized in \Sec{section_conclusion}.

\section{Electron signals from the Migdal effect}\label{section_migdal_effect}
In a bound-state atomic system, when the nucleus of the atom obtains a recoil momentum from a DM-nucleus scattering process, the electron cloud of the atom does not always follow the motion of the nucleus instantly. The momentum suddenly transferred into the nucleus can lead to the ionization or excitation of the atom, which is known as the Migdal effect.
In this work, we closely follow the formalism adopted in~\REF{Ibe:2017yqa} to calculate the electron energy spectrum from the Migdal effect. 
It is assumed that in DM-nucleus scattering process, in the first step, all the momentum is transferred into the nucleus within the atom. Then, the electrons of the atom are excited from ground states into continuum states due to the sudden perturbation from the recoil nucleus.
In the laboratory frame where the target atom is at rest, for a given momentum transfer $q$ from the DM particle to the nucleus,   the minimally required velocity of the DM particle  for a 
single-electron ionization with total energy $E_{\text{em}}$ is given by
\begin{align}
    v_{\min}\lra{q}=\frac{q}{2\mu_{\chi N}}+\frac{E_\text{em}}{q},\label{eq_vmin}
\end{align}
where $\mu_{\chi N}$ is the reduced mass of a DM particle and a nucleus $N$.
The total energy transferred into the electron cloud  $E_{\text{em}}$ can be written as
$E_\text{em}=E_{nl}+T_e$ with $T_e$  the kinetic energy of the ionized electron and $E_{nl}$  the binding energy of the initial state of the ionized electron with the primary quantum number $n$ and orbital quantum number $l$.

The differential cross section of the Migdal effect as a function of  $T_e$ and nuclear recoil energy $T_N$  can be approximately factorized into the product of elastic DM-nucleus scattering cross section ${d\sigma_{\chi N}}/{dT_N}$ and ionization probability ${dP_{nl}}/{d\ln T_e}$ as follows~\cite{Ibe:2017yqa, Essig:2019xkx} 
\begin{align} \label{eq_cs_mig0}
    \frac{d\sigma_{{\text{Mig}},nl}}{d T_Nd\ln T_e}
    \approx
    \frac{1}{2\pi}
    \frac{d\sigma_{\chi N}}{dT_N}
    \frac{dP_{nl}}{d\ln T_e}\lra{T_e,\,q_e} ,
\end{align}
where $q_e=m_e q/m_N$.  In the case of  $q_e r_A\ll 1$, where $r_A$ is the classical radius of the atom, the ionization probability ${dP_{nl}}/{d\ln T_e}$ can be related to the ionization factor of DM-electron scattering $\lrd{f_{nl}^{\text{ion}}\lra{k_e,\,q_e}}^2$ defined in Ref.~\cite{Essig:2019xkx},
\begin{align}
    \frac{dP_{nl}}{d\ln T_e}
    \approx
    \frac{\pi}{2}\lrd{f_{nl}^{\text{ion}}\lra{k_e,\,q_e}}^2,
\end{align}
where $k_e=\sqrt{2m_eT_e}$ is the momentum of the ionized electron.
Specifically, the ionization factor$\lrd{f_{nl}^{\text{ion}}\lra{k_e,\,q_e}}^2$ for closed-shell atoms like xenon can be calculated using non-relativistic quantum mechanical approaches as follows~\cite{Essig:2019xkx,Essig:2012yx,DarkSide:2018ppu,Essig:2017kqs},
\begin{align}
    \begin{split}
        \lrd{f_{nl}^{\text{ion}}\lra{k_e,\,q_e}}^2=&
        \frac{2k_e}{\pi}
        \sum_{l^{\prime}=0}^{\infty}\sum_{L=|l-l^{\prime}|}^{l+l^{\prime}}
        (2l^{\prime}+1)(2l+1)(2L+1)\\
        &\begin{pmatrix}l & l^{\prime} & L \\0 & 0 & 0\end{pmatrix}^{2}
        \left|\int dr r^2\widetilde{R}_{k_e l^{\prime}}^* j_{L}(q_er) {R}_{nl}\right|^{2},\label{eq_ionf_nonrel}
    \end{split}
\end{align}
where the big brackets stand for the Wigner-3j symbol, 
$j_L$ is the spherical Bessel function of order $L$,
$R_{nl}$ is the radial wave function of the bound state orbital,
and $\widetilde{R}_{k_e l^{\prime}}$ is the continuum wave function of electrons with momentum $k_e$.
Note that \eq{eq_ionf_nonrel} is only valid for closed-shell atoms like xenon, but is not necessarily the case with other atoms.
The normalization of the radial wave function in \eq{eq_ionf_nonrel} is
$\int drr^2R_{n^\prime l^\prime}^*R_{nl}=\delta_{n^\prime n}\delta_{l^\prime l}$
for bound states and 
$\int dr r^2\widetilde{R}_{k_e^\prime l^\prime}^*\widetilde{R}_{k_e l}=2\pi\delta\lra{k_e^\prime-k_e}\delta_{l^\prime l}$
for continuum states~\cite{Essig:2019xkx}.
In \REF{Essig:2012yx,Essig:2019xkx,DarkSide:2018ppu,Essig:2017kqs}, $R_{nl}$ is taken as the Roothaan-Hartree-Fock~(RHF) wave function~\cite{Bunge:1993jsz}, while $\widetilde{R}_{k_el^{\prime}}$ is determined by solving the Schr$\ddot{\text{o}}$dinger equation with central potential $Z_{\text{eff}}/r$ with $Z_{\text{eff}}=n\sqrt{E_{nl}/13.6~{\text{eV}}}$.
Since $\widetilde{R}_{k_el^{\prime}}$ determined in this approach is not orthogonal to $R_{nl}$, the $L=0$ term in \eq{eq_ionf_nonrel} has to be dropped out in order to restore the correct results.
In~\REF{Ibe:2017yqa}, the ionization factor ${dP_{nl}}/{d\ln T_e}$ is calculated using a fully relativistic method, where the continuum wave functions of electrons are orthogonal to the wave functions of bound state orbitals, and are all determined by solving the Dirac-Hartree-Fock equations with a relativistic self-consistent mean-field approach using the \texttt{Flexible Atomic Code}~(\texttt{FAC})~\cite{FAC, Ibe:2017yqa}. In this work, we adopt the numerical results of ${dP_{nl}}/{d\ln T_e}$ from~\cite{Ibe:2017yqa}. We also use the dipole approximation~\cite{Ibe:2017yqa,Essig:2019xkx} such that the dependence on $q_e$ of the ionization probability for $q_er_A\ll 1$ can be approximately written as
\begin{align}
    \frac{dP_{nl}}{d\ln T_e}\lra{T_e,\,q_e}
    \approx
    \lra{\frac{q_e}{q_0}}^2\frac{dP_{nl}}{d\ln T_e}\lra{T_e,\,q_0},
\end{align}
where $q_0$ is a reference momentum which also satisfies the condition $q_0r_A\ll 1$. Following~\REF{Ibe:2017yqa}, we take $q_0=1$~eV.
The binding energies for electron shells of xenon atoms, which are calculated in~\REF{Ibe:2017yqa} using the \texttt{FAC} code and in~\REF{Bunge:1993jsz} using the Roothaan-Hartree-Fock~(RHF) method, respectively, are shown in \tab{tab_biding_energy}.
For the Migdal effect of xenon atoms, we adopt the binding energies from the RHF methods since the binding energies calculated with the RHF methods are in better agreement with the current measurements related to the outer-shell electrons of xenon atoms~\cite{Baxter:2019pnz}.

For simplicity, we consider the spin-independent (SI) DM-nucleus scatterings through contact interactions, and also assume that the scattering is isospin-conserving.  In this scenario, the scattering cross section at the nucleus level $\sigma_{\chi N}$ is related to that at the nucleon level  $\sigma_{\chi p}$ as follows ~\cite{Lewin:1995rx},
\begin{align}
    \frac{d\sigma_{\chi N}}{dT_N}=
    \frac{F_N^2(q)}{T_N^{\max}}
    \frac{\mu_{\chi N}^2}{\mu_{\chi p}^2}A_N^2\sigma_{\chi p}
    ,\label{eq_cs_SI}
\end{align}
where $q^2=2m_N T_N$, 
$T_N^{\max}={4m_\chi m_N T_\chi}/{(m_\chi+m_N)^2}$ is
the maximal kinetic energy of the nucleus from the DM-nucleus elastic scatterings for a given initial kinetic energy $T_\chi$ of the DM particle, 
$\sigma_{\chi p}$ is the SI DM-nucleon cross section,
$A_N$ is the atomic mass number of a nucleus $N$,
and $F_N\lra{q}$ is the nuclear form factor
which is taken as the Helm form factor~\cite{Helm:1956zz,Lewin:1995rx},
\begin{align}
    F_N\lra{q}=\frac{3j_{1}(qR_{1})}{qR_{1}}e^{{-q^{2}s^{2}}/{2}},\label{eq_Helm}
\end{align}
where $j_{1}$ is the first-order spherical Bessel Functions,
$R_{A}=1.2A^{1/3}~{\text{fm}}$,
$R_{1}=\sqrt{R_{A}^{2}-5s^{2}}$,
and $s=1~{\text{fm}}$~\cite{Engel:1991wq}.

Finally, the electron event rate induced by the Migdal effect of halo DM at the surface of the Earth is given by
\begin{align}
    \frac{dR}{d E_\text{em}}=
    \sum_{nl}
    \frac{1}{m_N}
    \int_0^\infty
    d T_N
    \int_{v_{\min}\lra{q}}^\infty
    d^3 \boldsymbol{v}
    \frac{d\sigma_{{\text{Mig}}, nl}}{d T_N dT_e}v
    f_{\text{halo}}\lra{\boldsymbol{v}+\boldsymbol{v}_\oplus},\label{eq_mig_sp}
\end{align}
where $\boldsymbol{v}_\oplus$ is the velocity of the Earth in the galactic rest frame.
The velocity distribution of halo DM $f_{\text{halo}}(\boldsymbol{v})$ is assumed to be  the truncated Maxwell-Boltzmann distribution~\cite{Lewin:1995rx},
\begin{align}
    f_{\text{halo}}(\boldsymbol{v})
    =
    \frac{n_0}{N}\exp\lra{{-\frac{\boldsymbol{v}^2}{v^{2}_{0}}}}\Theta(v_{\text{esc}}-\lrd{\boldsymbol{v}}),\label{eq_halo}
\end{align}
where $N=(\pi v^{2}_{0})^{\frac{3}{2}}{\text{Erf}}(v_{\text{esc}}/v_{0})-2\pi v^{2}_{0}v_{\text{esc}}\exp\lra{{-v^2_{\text{esc}}/v^2_0}}$ with
${\text{Erf}}(x)$  the Gauss error function,
$v_0$ is the characteristic speed,
$v_{\text{esc}}$ is the galactic escape velocity,
and $\Theta$ is the Heaviside step function.
Note that $f_{\text{halo}}(\boldsymbol{v})$ defined in the above equation is normalized to the local number density $n_0$ with $n_0=\rho_0/m_\chi$ with $\rho_0$ is the local DM energy density at the surface of the Earth.

\begin{table}[tbh]\centering
    \begin{tabular}{@{}ccccccccc}
        \toprule
        Orbital binding energy & $5p^6$ & $5s^2$ & $4d^{10}$ & $4p^6$ & $4s^2$ & $3d^{10}$ & $3p^6$ & $3s^2$ \\
        \colrule
        RHF [eV] & 12.4 & 25.7 & 75.6 & 163.5 & 213.8 & 710.7 & 958.4 & 1093.2 \\
        \colrule
        FAC [eV] & 9.8 & 21 & 61 & 140 & 200 & 660 & 930 & 1100 \\
        \botrule
    \end{tabular}
    \caption{Binding energies of the electron shells of the xenon atom calculated from the \texttt{FAC} code~\cite{FAC,Ibe:2017yqa} and  the RHF method~\cite{Bunge:1993jsz}.}\label{tab_biding_energy}
\end{table}

\begin{figure}[tbh]
    \centering
    \includegraphics[width=\figsizeOne]{\plotsdir/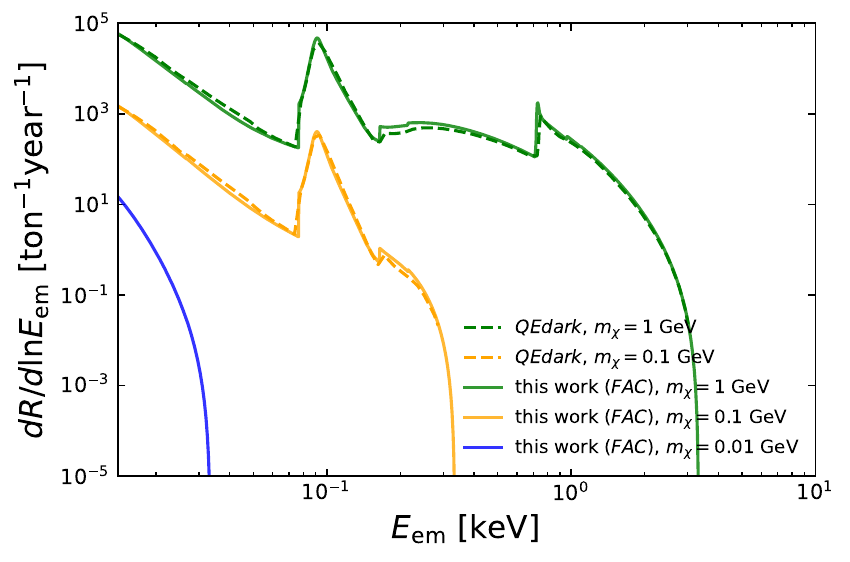}
    \caption{
        Electron event rates from the Migdal effect of xenon atoms at the surface of the Earth for different DM particle masses $m_\chi=0.01$~(blue solid), $0.1$~(orange solid), and $1~{\text{GeV}}$~(green solid), respectively.  The DM-nucleus scattering cross section is fixed as $\sigma_{\chi p}=10^{-38}~\text{cm}^2$, and the velocity of the Earth is taken as $v_\oplus=232~{\text{km/s}}$.
The results  calculated using the \texttt{QEdark} code~\cite{Essig:2019xkx} for $m_\chi=0.1$~(orange dashed) and $1~{\text{GeV}}$~(green dashed) are also shown for comparison.}
    \label{fig_mig_sp0}
\end{figure}

In \fig{fig_mig_sp0}, we show the electron event rates from the Migdal effect of the xenon atoms at the surface of the Earth for reference values of  
$\sigma_{\chi p}=10^{-38}~\text{cm}^2$ and $|\boldsymbol{v}_\oplus|=232~{\text{km/s}}$ with different DM masses $m_\chi=0.01$, $0.1$, and $1~{\text{GeV}}$, respectively. 
In the calculation, we take into account the contributions of the Migdal effect from the shells with  $n=3,4,5$ of the xenon atoms.
It should be emphasized that at low energies (typically $E_\text{em}\lesssim 75$~eV), the dominant ionization signals arise from the valence electrons. Due to the distortion of electron orbits resulting from the surrounding atoms in liquid xenon,
the ionization spectra from the $n=5$ shell may not be very accurate.
We use the benchmark parameters $\rho_0=0.3~{\text{GeV}/\text{cm}^3}$, $v_{\text{esc}}=544~{\text{km/s}}$, and $v_0=220~{\text{km/s}}$.
Due to the orbital motion of the Earth around the Sun, the value of $|\boldsymbol{v}_\oplus|$ varies slowly during a year. For instance, $|\boldsymbol{v}_\oplus|$ changes in the range  $219-248~{\text{km/s}}$ during the year 2022. A detailed discussion on the time dependence of $|\boldsymbol{v}_\oplus|$ can be found in \App{appendix_sidereal}.
It can be seen from the figure that for GeV scale halo DM particles, the Migdal induced electron energy can reach $\mathcal{O}(\text{keV})$, which can be probed by the low-threshold DD experiments.
For comparison purpose, the results from the \texttt{QEdark} code~\cite{Essig:2019xkx} are also shown in \fig{fig_mig_sp0}, which agree well with ours. We also cross-checked our results with that from the \texttt{wimprates} code~\cite{jelle_aalbers_2022_7041453} and that of XENON1T collaboration~\cite{Aprile:2019jmx} by substituting binding energies from the RHF methods with those from the $\texttt{FAC}$ code, and found their results were in perfect agreement with ours.

Recently, \REF{Xu:2023wev}
has reported the null result for the research on S1-S2 correlated signals induced by the Migdal effect in xenon time projection chamber~(TPC).
However, this null result does not mean that the Migdal effect does not exist due to the possible inaccuracy in the predictions for either the event rate of the Migdal effect or the signal response in liquid xenon.

\section{Diurnal modulation of electron events from the Migdal effect}\label{section_diurnal_effect}
\subsection{Isodetection angles}\label{subsection_isodetection_ring}
 
\begin{figure}[t]
    \centering
    \includegraphics[width=\figsizeOne]{\plotsdir/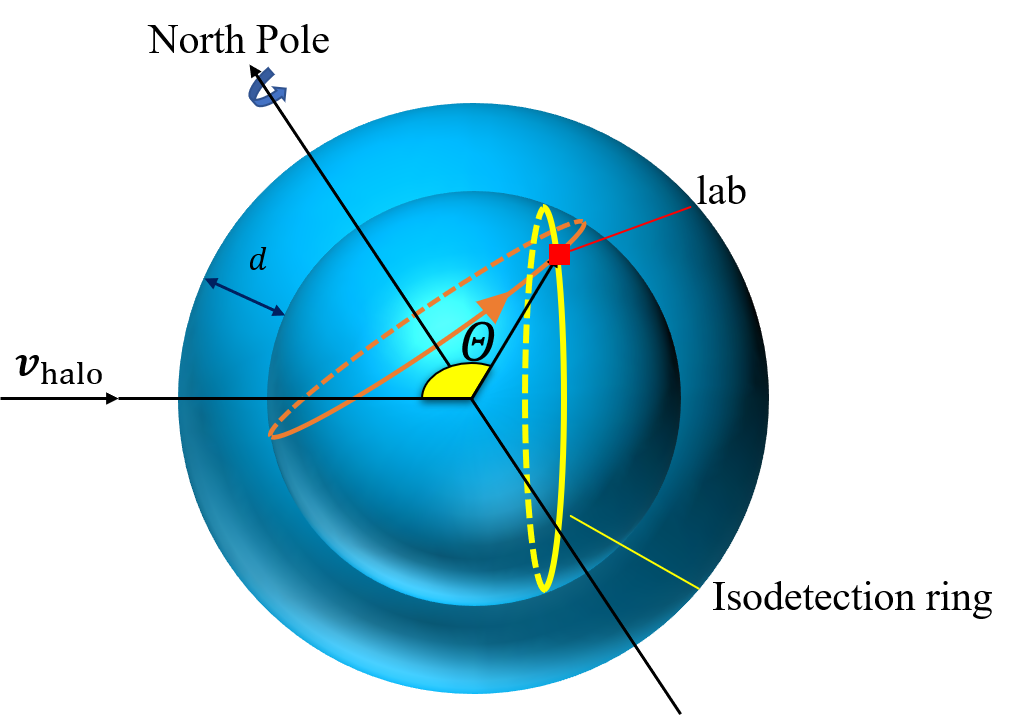}
    \caption{
Sketch of an isodetection ring with isodetection angle $\Theta$ and the trajectory of a laboratory at a depth $d$.
$\boldsymbol{v}_{\rm halo}$ is the overall velocity of DM halo in the rest frame of the Earth.
    }
    \label{fig_isodetection_ring}
\end{figure}

Before reaching the underground detectors, DM particles can be deflected and lose energy due to the scatterings off the nuclei within the Earth. For sufficiently large DM-nucleus scattering cross sections, the underground DM fluxes are expected to be significantly different from that at the surface of the Earth.
Due to the motion of the Earth relative to the Galaxy,
the DM halo has an overall velocity $\boldsymbol{v}_{{\text{halo}}}=-\boldsymbol{v}_\oplus$ in the rest frame of the Earth~(see, \fig{fig_isodetection_ring}),
which suggests that the DM flux at the surface of the Earth is not isotropic but has an azimuthal symmetry around the direction of $\boldsymbol{v}_{{\text{halo}}}$.
Since the matter density distribution of the Earth can be well approximated as layers with spherical symmetry around the center of the Earth~\cite{Dziewonski:1981xy}, it is expected that the underground DM velocity distribution should have the same azimuthal symmetry as well. 
It is convenient to use the concepts of isodetection angle (i.e. the polar angle)
and isodetection ring for diurnal modulation analysis~\cite{Collar:1992qc, Emken:2017qmp}.
We follow the convention of \REF{Emken:2017qmp} to define the isodetection angle $\Theta$ as the angle between the position vector of the laboratory relative to the center of the Earth and the inverse direction of $\boldsymbol{v}_{{\text{halo}}}$. The positions with the same isodetection angle $\Theta$ at a depth $d$ form an isodetection ring. The isodetection angle and isodetection ring are illustrated in \fig{fig_isodetection_ring}.
The DM fluxes and therefore the event rates measured by the DD experiments on the same isodetection ring are expected to be the same due to the azimuthal symmetry.
Since the spin axis of the Earth is not parallel to $\boldsymbol{v}_{{\text{halo}}}$, an underground laboratory will pass through different isodetection rings during a day as shown in \fig{fig_isodetection_ring}, which leads to the modulation of the recoil event rate observed by the experiment.
Once the underground DM fluxes at different isodetection rings are obtained, the time dependence of the event rate for a given experiment can be calculated in a straightforward way.

In the rest frame of the Earth,  the direction of $\boldsymbol{v}_{\text{halo}}$  can be described by the declination $\delta_{\text{halo}}$ and right ascension $\alpha_{\text{halo}}$ angles in the equatorial coordinate. The isodetection angle for an underground laboratory (at longitude  $\lambda_{\text{lab}}$, latitude $\varphi_{\text{lab}}$, and with a depth $d$) at a given  time $t$ in Universal Time~(UT)  can be expressed as follows
\begin{align}
    \begin{split}
        \cos{\Theta}
        =
        -
        \sin{\delta_{\text{halo}}}\sin{\varphi_{\text{lab}}}
        -
        \cos{\delta_{\text{halo}}}\cos{\varphi_{\text{lab}}}
        \cos\phi\lra{t},
        \label{eq_isodetection_angle_time}
    \end{split}
\end{align}
where 
$\phi$ is a time-dependent phase.
The temporal evolution of $\phi(t)$ is related to the local sidereal time $\tau(t,\lambda_{\text{lab}})$ as
\begin{align}
    \phi(t)=\frac{2\pi}{T_\tau} \tau\lra{t,\,\lambda_{\text{lab}}}-\alpha_{\text{halo}},
\end{align}
where $T_\tau=86164.1$~s~\cite{ParticleDataGroup:2022pth} is the length of a sidereal day defined as the period of the self-rotation of the Earth, which is 24 sidereal hours.  

The local sidereal time $\tau$ is a function of  $t$ and $\lambda_{\text{lab}}$. The details of the local sidereal time can be found in the \App{appendix_sidereal}.
The minimal and the maximal isodetection angle for a given $\delta_{\text{halo}}$ and $\varphi_{\text{lab}}$ during a sidereal day are given by
$
\Theta_{\min} = \lrd{\varphi_{\text{lab}}+\delta_{\text{halo}}}
$
and
$
\Theta_{\max} = \pi - \lrd{\varphi_{\text{lab}}-\delta_{\text{halo}}}
$,
respectively.
The direction of $\boldsymbol{v}_{\text{halo}}$ varies slightly during a year due to the revolution of the Earth around the Sun.
As an example, in the year 2022,
the value of $\delta_{\text{halo}}$ takes the maximal~(minimal) value $-41^\circ~(-55^\circ)$ on the date $t_{a(b)}=20\text{th},\,{\text{Apr.}},\,2022~{\text{UT}}~(30\text{th},\,{\text{Oct.}},\,2022~{\text{UT}})$,
and $\alpha_{\text{halo}}$ takes the minimal~(maximal) value $123^\circ~(144^\circ)$ on the date $t_{c(d)}=21\text{st},\,{\text{Jan.}},\,2022~{\text{UT}}~(14\text{th},\,{\text{Aug.}},\,2022~{\text{UT}})$. 
Details on the variation of $\delta_{\text{halo}}$ and $\alpha_{\text{halo}}$ during the year 2022 can be found in \App{appendix_sidereal}.
In \fig{fig_isodetection_angle_sidereal_time},
we show the variation of the isodetection angles of the following two laboratories:
the Gran Sasso National Laboratory of INFN~(\Gran) at $42.4^{\circ}$~N,\,$13.5^{\circ}$~E with a depth $d=1.4~{\text{km}}$ and the China Jinping Underground Laboratory~(\Jinp) at $28.2^{\circ}$~N,\,$101.7^{\circ}$~E with a depth $d=2.4~{\text{km}}$ during a sidereal day.
The zero-hour of the sidereal time is defined as when the vernal equinox comes across the local meridian.
\begin{figure}[t]
    \centering
    \includegraphics[width=\figsizeTwo]{\plotsdir/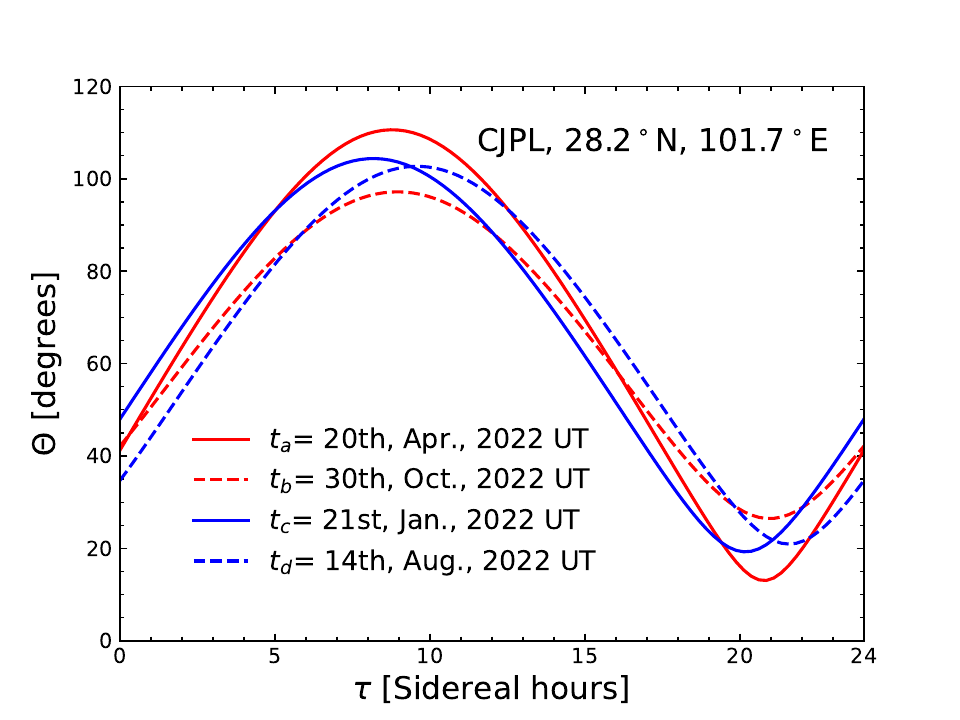}
    \includegraphics[width=\figsizeTwo]{\plotsdir/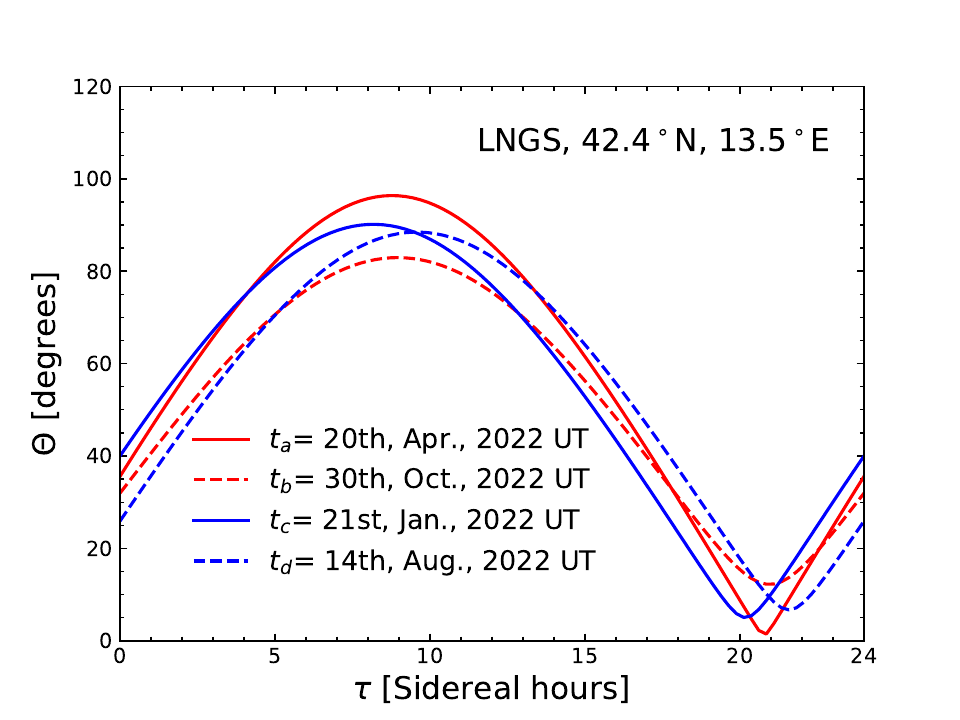}
    \caption{
        The variation of the isodetection angles 
        during a sidereal day
        on the four typical dates $t_a$, $t_b$, $t_c$, and $t_d$ at the two underground laboratories \Jinp~(left panel) and \Gran~(right panel).}
    \label{fig_isodetection_angle_sidereal_time}
\end{figure}
As shown in \fig{fig_isodetection_angle_sidereal_time},
the difference between the maximal and minimal isodetection angle in a sidereal day takes the maximal~(minimal) value of the year 2022 on the date $t_{a(b)}$,
which makes the variation of the diurnal modulation signal the strongest~(weakest) one of the year 2022. 
The time phase of the peak of the isodetection angle in a sidereal day becomes the earliest~(latest) one of the year 2022 on the date $t_{c(d)}$,
which makes the time phase of the diurnal modulation signal the earliest~(latest) one of the year 2022.

\subsection{Underground DM distributions}\label{subsection_Earth_attenuation}
In the calculation of the DM-nucleus scattering within the  Earth, for simplicity, we only consider the elastic DM-nucleus scattering which is an irreducible process for DD experiments.
For sub-GeV halo DM, since the factor $qs\lesssim 2\mu_{\chi N}v_{\text{esc}}s\approx 1.8\times 10^{-5}\times{m_\chi}/\text{MeV} \ll 1$ in the expression of the form factor~\eq{eq_Helm}, the value of the nuclear form factor $F_N\lra{q}$ is very close to unity. Thus, the effect of the nuclear form factor for halo DM can be safely neglected.
Note that for CR boosted DM the form factor may play a significant role~\cite{Xia:2021vbz,Super-Kamiokande:2022ncz}.

For the process of DM-nucleus elastic scattering, in the laboratory frame where the target nucleus is at rest, the nuclear recoil energy $T_N$ can be written as $T_N={T_N^{\max}}\lra{1-\cos{\theta_\chi^*}}/2$,
where $T_N^{\max}$ is the same as that in \eq{eq_cs_SI} and $\theta_\chi^*$ is the scattering angle of the DM particle in the center-of-mass frame of the DM particle and the nucleus.
The relation between the scattering angle $\theta_\chi$ of the non-relativistic DM particle in the laboratory from  $\theta_\chi^*$ is given by
\begin{align}
    \cos{\theta_\chi}=\frac{m_\chi+m_N\cos{\theta_\chi^*}}{\sqrt{m_\chi^2+m_N^2+2m_\chi m_N\cos{\theta_\chi^*}}}.\label{eq_scattering_angle}
\end{align}
The Earth shielding effect has been estimated using a simplified analytical method based on the ballistic approximation~\cite{Kouvaris:2014lpa, Starkman:1990nj, Kavanagh:2017cru, Bringmann:2018cvk, Xia:2020wcp, Ge:2020yuf}.
In this approach,
the decrease of the DM kinetic energy $T_\chi$ with respect to the propagation distance $z$ of the DM particle  due to the elastic scatterings off the nuclei within the Earth is given by
\begin{align}
    \frac{d T_{\chi}}{d z} = -\sum_{N}^{ }n_{N}\int_{0}^{T_N^{\max}}\frac{d\sigma_{\chi N}}{d T_{N}}T_{N}d T_{N},
    \label{eq_elf}
\end{align}
where $n_N$ is the number density of element $N$ in the Earth and $d\sigma_{\chi N}/d T_{N}$ is given by \eq{eq_cs_SI}.
In the ballistic approximation, it is assumed that the DM particles travel in straight lines and only the average energy loss is considered. However, this ballistic approximation is not appropriate for the calculation of the Earth shielding effect of sub-GeV halo DM.
For sub-GeV halo DM particles whose masses are much lower than the nuclear mass, the deflection of the DM particle at each scattering can be very significant. This can be seen from \eq{eq_scattering_angle}.  When $m_\chi \ll m_N$, the scattering angle $\theta_\chi$ is approximately equal to $\theta_\chi^*$, which implies that $\theta_\chi$ can be large since the scattering process described by \eq{eq_cs_SI} is isotropic in the center-of-mass frame.
The deflection makes the trajectories of the DM particles much longer than that in the ballistic approximation, which could result in significant changes of  DM phase space distribution at deep underground.

A more realistic description of the Earth shielding effect can be obtained by numerical Monte-Carlo (MC) simulation~\cite{Xia:2021vbz, Cappiello:2019qsw, Emken:2018run, CDEX:2021cll, PROSPECT:2021awi}. In this work, the Earth shielding effect of sub-GeV halo DM is simulated using the \texttt{DarkProp} code~\cite{Xia:2021vbz},
which is a general purpose package which supports  both relativistic and non-relativistic DM,
takes into account the geometry of the Earth, and has the flexibility to customize the DM initial velocity distribution and scattering cross section.
There are also other publicly available codes developed to calculate the Earth shielding effect for specific models~\cite{Bringmann:2018lay, Kavanagh:2016pyr, Kavanagh:2017cru, Emken:2018run, Emken:2017qmp, Emken:2021lgc, Bramante:2022pmn, Chen:2021ifo,Cappiello:2023hza}.
In the numerical simulation, we sample the initial velocities of halo DM particles at the surface of the Earth according to the standard halo model velocity distribution in the rest frame of the Earth $f_{\text{halo}}\lra{\boldsymbol{v}+\boldsymbol{v}_\oplus}$ given by \eq{eq_halo}.
The simulation of a DM particle trajectory is divided into free-propagation and scattering processes.
For the typical kinetic energy of DM particles, the gravity of the Earth can be safely neglected so that the free-propagation is along straight lines. The free-propagation length is sampled according to the mean-free-path calculated from the total cross section by integrating \eq{eq_cs_SI} and the number densities of the nuclei within the Earth.
After free-propagation, the DM particle scatters with a nucleus sampled according to the chemical abundance and the corresponding total cross section, and the scattering angle is sampled according to the differential cross section of \eq{eq_cs_SI}.
The free-propagation and scattering processes repeat until the DM particle leaves the surface of the Earth or its kinetic energy drops below the detection threshold of DD experiments. We take this threshold of DM speeds as $30~{\text{km/s}}$ in this work.
We record the momentum and position of each DM particle when it comes across the surface of the sphere at a given depth and isodetection ring.
Based on the ensemble of the recorded events, we reconstruct the underground speed distributions $f\lra{v,\,d,\,\Theta}$ which is normalized to the local underground DM number density, for $v>30~{\text{km/s}}$ on each isodetection ring at a depth $d$.

\subsection{Diurnal modulation amplitudes}

\begin{figure}[t]
    \centering
    \includegraphics[width=\figsizeTwo]
{\plotsdir/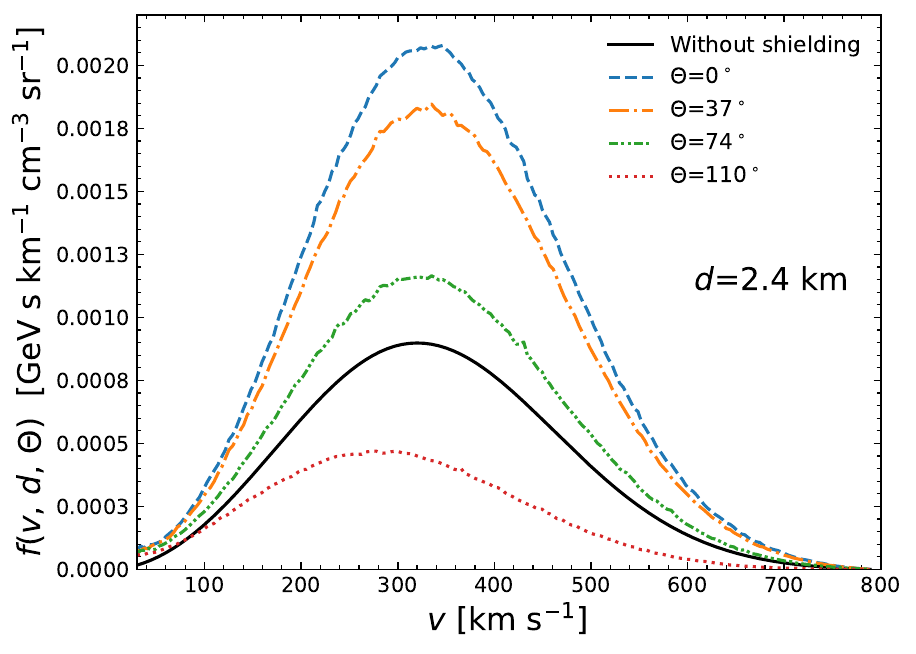}
    \includegraphics[width=7.6cm]
    {\plotsdir/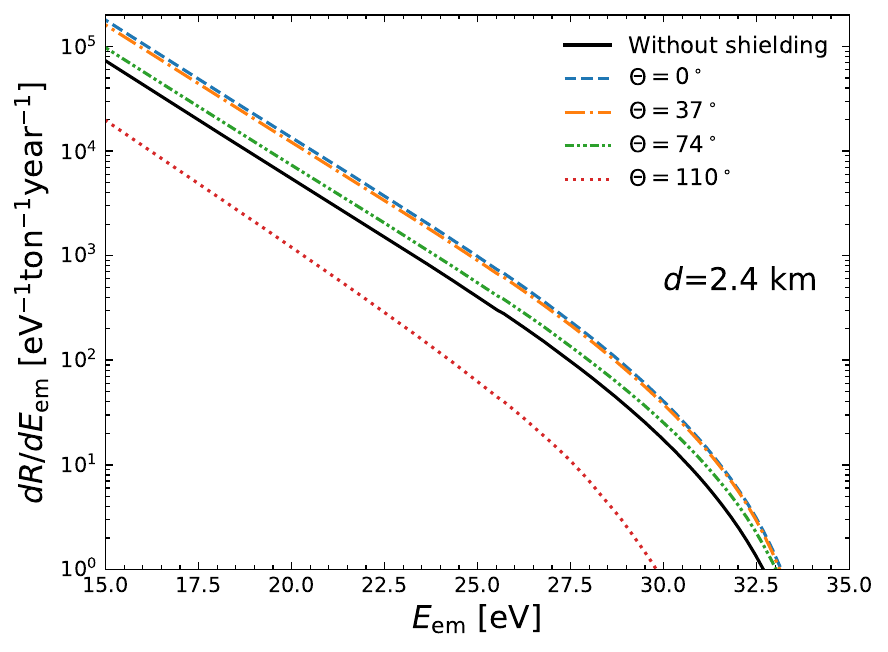}
    \caption{
(Left)
        Underground DM speed distributions at a depth of 2.4~km on the date $t_a$
for a selection of isodetection rings with $\Theta = 0^\circ$, $37^\circ$, $74^\circ$, and $110^\circ$.
The DM particle mass and DM-nucleon scattering cross section are taken from the benchmark values in  \eq{eq:typical_paras}. 
The black solid curve indicates the DM speed distribution without the Earth shielding effect.
(Right)
        Electron event rates from the Migdal effect of xenon atoms corresponding to the underground DM speed distributions on the left panel.  The electron event rate without the Earth shielding effect is also shown as the black solid curve for comparison.
        }
    \label{fig_fv_sp}
\end{figure}

On the left panel of~\fig{fig_fv_sp},
we show the underground speed distributions which are normalized to the underground DM density, at the depth $d=2.4~\text{km}$ on different isodetection rings on the date $t_a$.
In this work, we focus on the following benchmark values of DM particle mass and DM-nucleon scattering cross sections
\begin{equation}
    m_\chi=10~\text{MeV}~ \text{and}~ \sigma_{\chi p}=10^{-33}~\text{cm}^2,
    \label{eq:typical_paras}
\end{equation}
which are slightly below the current experimental upper limits.
In the MC simulations, the density profile of the Earth is taken from the preliminary reference Earth model~(PREM)~\cite{Dziewonski:1981xy}. The chemical abundances of the core and the mantle of the Earth are taken from~\cite{McDonough:2003}.
We divide the sphere at the depth $d=2.4~{\text{km}}$ into 180 equal-area isodetection rings and simulate $\mathcal{O}(10^7)$ events passing through each isodetection ring.
As shown on the left panel of \fig{fig_fv_sp}, the underground speed distributions increase as $\Theta$ decreases.
For $\Theta \lesssim 85^\circ$, the underground speed distributions are higher than that without the Earth shielding effect.
The mean-free-path is about $377~{\text{km}}$ with benchmark parameters given by \eq{eq:typical_paras} in consideration of the homogeneous Earth model we discussed in \Sec{section_introduction}, which is far longer than the depth $d=2.4~{\text{km}}$ but far shorter than the diameter of the Earth which is about $\mathcal{O}(10^4)~{\text{km}}$.
This value of the mean-free-path indicates that the DM particles outside the Earth can easily pass through the crust of the Earth to reach the depth $d$ but are almost blocked if they are injected from the back of the Earth.
For halo DM whose velocity distribution has a preferred direction along the halo velocity, the DM particles from the preferred direction can be almost directly injected onto the isodetection ring with small $\Theta$~(e.g. $\Theta=0$) but can hardly reach that with large $\Theta$~(e.g. $\Theta=\pi$) on the other side of the Earth due to multiple scatterings.
Moreover, the DM particles that are injected into the Earth can be deflected by the nuclei in the Earth.
Therefore, in a given time interval, the isodetection ring with small $\Theta$ can not only receive DM particles that are directly injected from the preferred direction but also DM particles that were injected before but have been deflected back, leading to the enhancement of underground DM distributions.
Our results are consistent with that obtained from the \texttt{DaMaSCUS} code~\cite{Emken:2017qmp}.
More details of the speed distribution reconstruction process can be found in \App{appendix_mc_details}.

The electron event rate of the Migdal effect at a given depth $d$ and isodetection angle $\Theta$ is given by
\begin{align}
    \frac{dR}{d E_\text{em}}=
    \sum_{nl}
    \frac{1}{m_N}
    \int_0^\infty
    d T_N
    \int_{v_{\min}}^\infty d v
    \frac{d\sigma_{{\text{Mig}}, nl}}{d T_N dT_e}v
    f\lra{v,\,d,\,\Theta}.
    \label{eq_mig_sp_d}
\end{align}
On the right panel of~\fig{fig_fv_sp},
we show the electron event rate of the Migdal effect of xenon atoms from the underground DM speed distributions at the depth $d=2.4~\text{km}$ for different isodetection angles on the date $t_a$.
As shown on the right panel of \fig{fig_fv_sp}, the electron event rates of the Migdal effect decrease as $\Theta$ increases.
The electron event rates of the Migdal effect with $\Theta \lesssim 85^\circ$ are also higher than that without the Earth shielding effect due to the enhancement of underground DM distributions resulting from the deflections.

The total electron event rate in the energy range $(E_1, E_2)$ at a given time $t$ and depth $d$ is given by $R(t,d)=\int^{E_2}_{E_1} dE_\text{em} dR/dE_\text{em}$.
To describe the modulation strength of the diurnal effect, we introduce a diurnal modulation amplitude in a sidereal day from time $t_0$ to $t_0+T_\tau$ as follows
\begin{align}
    A(t)\equiv
    \frac{
R(t,d)
    }{\frac{1}{T_\tau} \int_{t_0}^{t_0+T_\tau} d t  R \lra{t, d}} .
\end{align}
It is evident that $A(t)$ approaches unity when the time-variation in $R(t,d)$ disappears.

\begin{figure}[t]
\includegraphics[width=\figsizeTwo]{\plotsdir/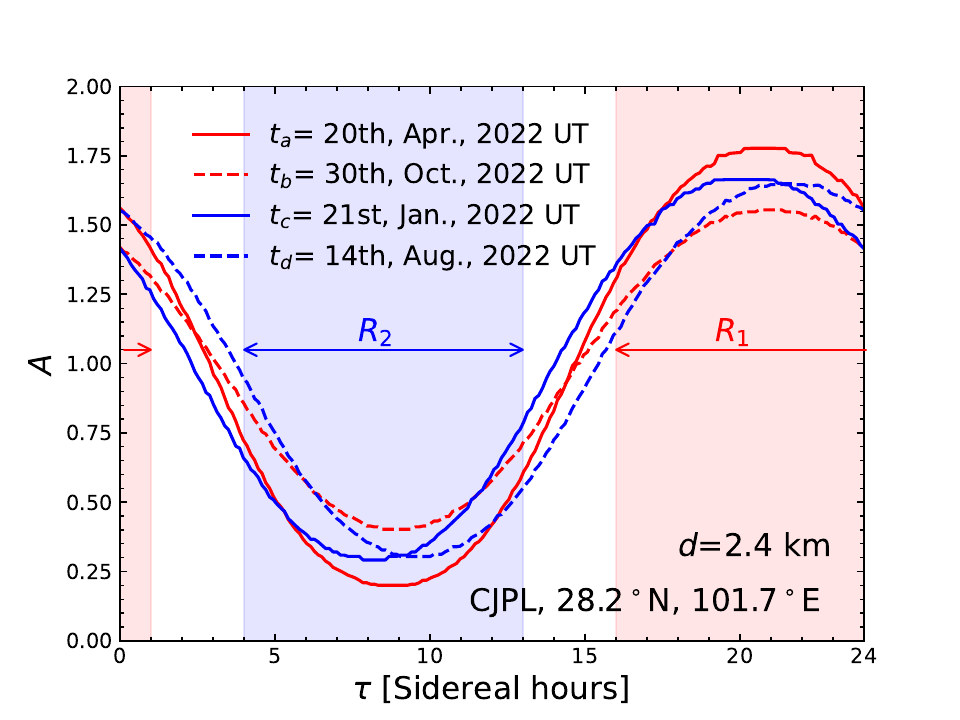}
\includegraphics[width=\figsizeTwo]{\plotsdir/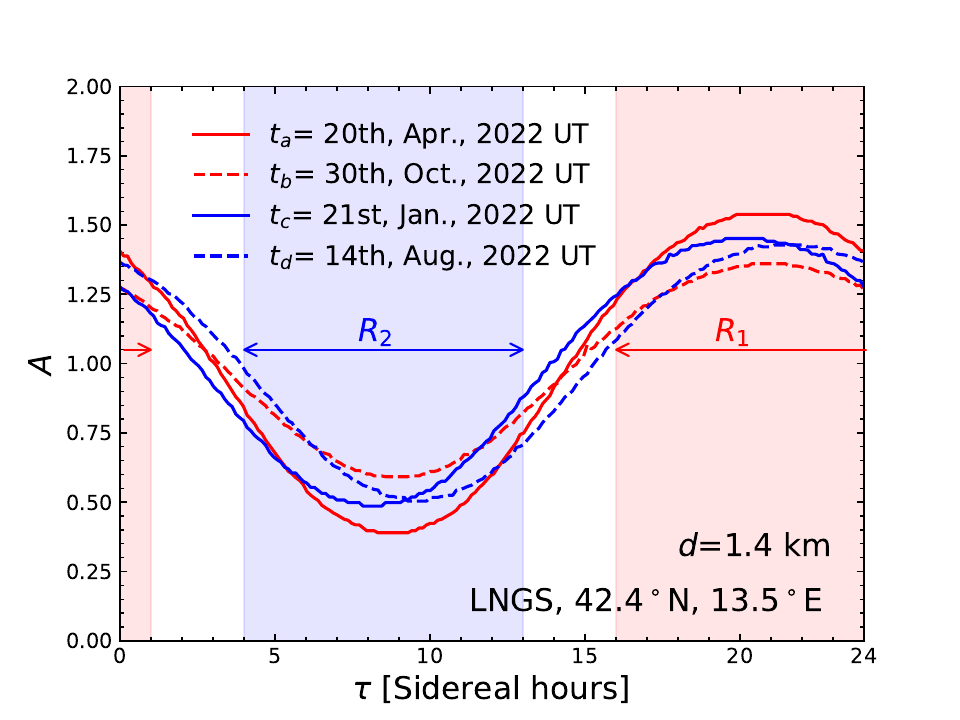}
    \caption{
        (Left)
Diurnal amplitude $A(t)$ of the electron events from the Migdal effect of xenon atoms as a function of the sidereal time at \Jinp~with a depth of 2.4~km on the four typical dates $t_a$, $t_b$, $t_c$, and $t_d$.
The DM particle mass and DM-nucleon scattering cross section are taken from the benchmark values in  \eq{eq:typical_paras}. 
The two equal-length sidereal time intervals $R_1$~(red shaded) and $R_2$~(blue shaded) chosen for evaluating the event-number asymmetry are also shown.
(Right)
The same as the left but for \Gran~with a depth of 1.4~km.
    }
    \label{fig_diurnal_effect_amp_PREM}
\end{figure}

In~\fig{fig_diurnal_effect_amp_PREM},
we show the diurnal modulation amplitudes from the Migdal effect of xenon atoms at \Jinp~and \Gran~on the four typical dates $t_{a,b,c,d}$ shown in \fig{fig_isodetection_angle_sidereal_time} for the benchmark parameters in \eq{eq:typical_paras}.
The electron event rate is calculated from the threshold of the Migdal ionization determined by the binding energy for $5p$ shell $E_{5p}=12.4~\text{eV}$ of xenon atoms to the maximal value of $E_{\text{em}}^{\max}=1/2\mu_{\chi N} v_{\max}^2 \simeq 33.5~\text{eV}$ allowed by the kinematics,
where
$v_{\max}=v_\text{esc}+v_\oplus$ with 
the velocity of the Earth and galactic escape velocity in the galactic rest frame taken as
$v_\oplus=232~\text{km}/\text{s}$ and $v_\text{esc}=544~\text{km}/\text{s}$, respectively.
The variation of the diurnal modulation amplitudes on a given date depends on the latitudes of the laboratories which determine the isodetection rings that the laboratories pass through.
For the date $t_{a(b)}$, the variation of the diurnal amplitudes during a sidereal day is the strongest~(weakest) of the year 2022.
The maximal and the minimal diurnal amplitudes can reach  $\sim 180\,\%$~($155\,\%$) and $\sim 20\,\%$~($40\,\%$), respectively, for \Jinp~(\Gran) on the date $t_a$.
The diurnal modulation amplitudes at \Jinp~have stronger variation than those at \Gran~during a sidereal day due to their different latitudes.
The time phases of the diurnal amplitudes depend on the right ascension $\alpha_{\text{halo}}$ of $\boldsymbol{v}_{\text{halo}}$ in the equatorial coordinate, which is determined for a given date.
For the date $t_{c(d)}$, the time phase of the diurnal modulation signal gets the earliest~(latest) of the year 2022.
The difference between the time phase on the date $t_{c}$ and $t_{d}$ is about $1.4$ sidereal hours.

\section{New constraints from PandaX-II and PandaX-4T}\label{section_constraints}

The Migdal effect has been considered to constrain sub-GeV DM by 
the experiments with dual-phase time projection chambers~(TPCs) which utilize the ionization electron data  (S2-only) to lower the detection thresholds~\cite{Aprile:2019jmx,Essig:2019xkx,DarkSide:2022dhx}.
Recently, the PandaX-II (PandaX-4T) experiment has released the S2-only data in the low photoelectron (PE) range of 50-75~PE (60-200~PE)  with an effective exposure of  $0.13~\text{ton}\cdot\text{year}$ ($0.55~\text{ton}\cdot\text{year}$) \cite{Cheng:2021fqb,PandaX:2022xqx}. 
The exposure of PandaX-4T is much larger than that obtained by previous experiments, but so far only the S2-only signals above 60~PE were reported, which makes the data less sensitive to light DM particles below $\sim 20$~MeV.  In this work, we derive the constraints on light DM from the PandaX-II and PandaX-4T data using the Migdal effect and compare them with that from XENON10 and XENON1T. The result shows that PandaX-II/4T are currently placing the leading constraints on DM-nucleon scattering cross section for sub-GeV DM particles.

\subsection{Data analysis procedures}
We first calculate the S2 signals induced by the Migdal effect for the PandaX-II experiment.  
For the data analysis of electron recoil signals, we closely follow the procedures in ~Refs.~\cite{Cheng:2021fqb, PandaX-II:2020oim, PandaX-II:2021jmq, Essig:2017kqs}.
For a given energy $E_{\text{em}}$ deposited into the xenon detector, 
the number of electron quanta $n_e$ is simulated using a normal distribution
\begin{align}
    n_e \sim \text{norm}\lra{n_e^0,\,f_l\cdot n_e^0} ,
\end{align} 
where $n_e^0=(1-r)+Q_y E_{\text{em}}$ is the mean value of $n_e$ with charge yield $Q_y=61~\text{keV}^{-1}$ taken from the constant charge yield model from~\cite{Cheng:2021fqb},
$\,f_l\cdot n_e^0$ is the standard deviation of $n_e$,
and $r$ is the recombination probability of the primary ionized electron.
Note that the value of $r$ almost vanishes at low kinetic energies for electron recoils~\cite{Sorensen:2011bd, Essig:2012yx, Essig:2017kqs}. 
The function $f_l$ describes the variation of $n_e$ which is in general energy dependent~\cite{PandaX-II:2021jmq}.
The ionized electrons can be drifted into the surface between liquid and gaseous xenon due to the strong electric field. The number of the electron quanta $N^\prime_e$  drifted into the surface is generated from a binomial distribution
\begin{align}
    N'_e\sim  \text{binom}(n_e, s) ,
\end{align}
where $s$ is  the survival probability under the  drift electric field
\cite{PandaX-II:2021jmq}.
The number of the electron quanta $N_e^{\prime\prime}$ extracted into xenon gas is generated by another binomial distribution
\begin{align}
    N_e^{\prime\prime}\sim  \text{binom}(N_e^\prime,\,\text{EEE}) ,
\end{align}
where $\text{EEE}$ stands for the electron extraction efficiency~\cite{Cheng:2021fqb}.
The raw S2 scintillation signals produced in gaseous xenon ${\text{S2}_\text{raw}}$ are generated by a normal distribution 
\begin{align}
    {\text{S2}_\text{raw}}\sim 
    {\text{norm}}(\,N_e^{\prime\prime}\cdot\text{SEG}, \sigma_\text{SE}\sqrt{N_e^{\prime\prime}}),
\end{align}
where $\text{SEG}$ and $\sigma_\text{SE}$ stand for the single-electron gain and its resolution~\cite{Cheng:2021fqb}, respectively.
The $\text{S2}_\text{raw}$ signals generated in the xenon gas can be captured by the photomultiplier tubes~(PMTs) as the $\text{S2}$ signals with an efficiency which is taken from~\cite{Cheng:2021fqb}.
In addition, due to the nonlinear effect of the baseline suppression firmware~\cite{PandaX-II:2020oim, PandaX-II:2021jmq}, the final S2 signals are given by ${\text{S2}}\cdot f_2\lra{\text{S2}}$, where the value of the S2-dependent function $f_2(\text{S2})$ is taken from~\cite{PandaX-II:2020oim}.
Note that the value of the function $f_l$, which describes the variation of $n_e$, for $T_e \lesssim 1~{\text{keV}}$ and the mean value of the survival probability $s$ were not explicitly published by the PandaX-II collaboration. 
The mean value of $s$ should lie somewhere between 0.6 and 1.0 according to~\cite{PandaX-II:2021jmq}.
For simplicity, we treat both $f_l$ and $s$ as constants and tune their values through best reproducing the PandaX-II results on the DM-{\it electron} scattering cross section through contact interactions~\cite{Cheng:2021fqb}.
We find that taking $s \approx 0.8$ and $f_l \approx 0.1$ makes our results most consistent with that of PandaX-II~\cite{Cheng:2021fqb}. Therefore, we take $s=0.8$ and $f_l=0.1$ as the benchmark values for the analysis of the Migdal effect in this work. 
Details about the PandaX-II electron recoil data analysis can be found in Appendix~\ref{appendix_PandaX-II}.

For the data analysis of PandaX-4T, we adopt the same analysis framework of PandaX-II to calculate the S2 signals except for the parameters of the EEE, SEG, $\sigma_\text{SE}$, and detection efficiency. 
We instead take $\text{EEE}=0.926, \text{SEG}=19.32$, $\sigma_\text{SE}/\text{SEG}=0.27$ from the PandaX-4T experiment in~\cite{PandaX-4T:2021bab, Zhang:2022wzy}.
The detection efficiency for PMTs of PandaX-4T is taken from~\cite{PandaX:2022xqx}.

For comparison purposes, we  independently analyzed the S2-only data of XENON10~\cite{Essig:2017kqs, Essig:2019xkx} and XENON1T~\cite{xenon1t:s2only_data_release, Aprile:2019xxb}.
The data analysis procedures of XENON10 are very similar to that of PandaX-II except for the number of electron quanta is alternatively defined as $n_e=n_e'+n_e''$, where $n_e'$ and $n_e''$ are generated by binomial distributions $n_e'\sim \text{binom}\lra{1,\,1-r}$ and $n_e''\sim\text{binom}\lra{\text{floor}\lra{E_\text{em}/W},\,f_e}$, respectively, and 
$f_e\simeq 0.83$ is the mean fraction of the total electron quanta $\text{floor}\lra{E_\text{em}/W}$
with $W=13.8$~eV  the average energy required to create a single quantum in liquid xenon~\cite{Essig:2012yx, Essig:2017kqs,Essig:2019xkx}.
For XENON1T, we follow the data analysis procedures provided by the publicly available code~\cite{xenon1t:s2only_data_release}, where the XENON1T collaboration has provided a response matrix including efficiency taken into account which can transform the spectra of $E_\text{em}$ into S2 signals by simple integration.
In \REF{Aprile:2019xxb}, the XENON1T collaboration has considered the cases with or without a   lower cutoff on $E_\text{em}$ at $0.186~{\text{keV}}$.
The result without the cutoff has been used to directly compare with the results from other experiments such as XENON10~\cite{Essig:2017kqs} and DarkSide-50~\cite{DarkSide:2018ppu}.
Therefore, for comparison purpose, we do not adopt the lower cutoff in performing the integration with respect to $E_\text{em}$.
More details on the data analysis procedures of XENON10 and XENON1T are summarized in \App{appendix_XENON10} and \App{appendix_XENON1T}, respectively.

\subsection{Constraints on DM-nucleon scattering cross sections}
We use MC methods to sample the DM-induced events  and obtain the S2 event rate 
$R_{\text{S2}}(t,d)$ in a given S2 bin which is, in general, a function of time $t$ and also depends on the depth $d$ of the laboratory due to the Earth shielding. 
The total event number in an S2 bin for an experiment taking data  from Universal Time $t_1$ to $t_2$ is given by
\begin{align} \label{eq_event_number}
    N_{\text{S2}}=
    \int_{t_1}^{t_2}
    dt
    R_{\text{S2}}
    \lra{t,\,d} . 
\end{align}
The predicted event numbers are compared with the data, and then the constraints on DM-nucleon scattering cross sections at $90\,\%$ C.L.  are obtained using the Binned Poisson method~\cite{Savage:2008er, Green:2001xy}.
We use the data ranging from 50 to 75~PE with 5 bins for PandaX-II~\cite{Cheng:2021fqb}, and 
60 to 200~PE with one bin  for PandaX-4T~\cite{PandaX:2022xqx}, respectively.
For XENON10, we use the  data from 14 to 203~PE with 7 bins~\cite{Essig:2017kqs} and from 150 to 3000~PE with12 bins for XENON1T~\cite{xenon1t:s2only_data_release}.
For PandaX-II and XENON10, we neglect the background rates as so far there is no estimation of the background available from these experiments.
For XENON1T, we take into account the electron recoil background and coherent neutrino-nucleus scattering background~($\text{CE}\nu\text{NS}$) from the code in Ref.~\cite{xenon1t:s2only_data_release}.
For PandaX-4T, in order to obtain conservative constraints, we take the prior background rate from~\cite{PandaX:2022xqx}.

In the first step, 
we derive the constraints on SI DM-nucleon scattering cross section through the Migdal effect using the S2-only data of PandaX-II~\cite{Cheng:2021fqb}, PandaX-4T~\cite{PandaX:2022xqx}, and XENON1T~\cite{Aprile:2019xxb,xenon1t:s2only_data_release} with the Earth shielding effect neglected.
\begin{figure}[t]
    \includegraphics[width=\figsizeTwo]{\plotsdir/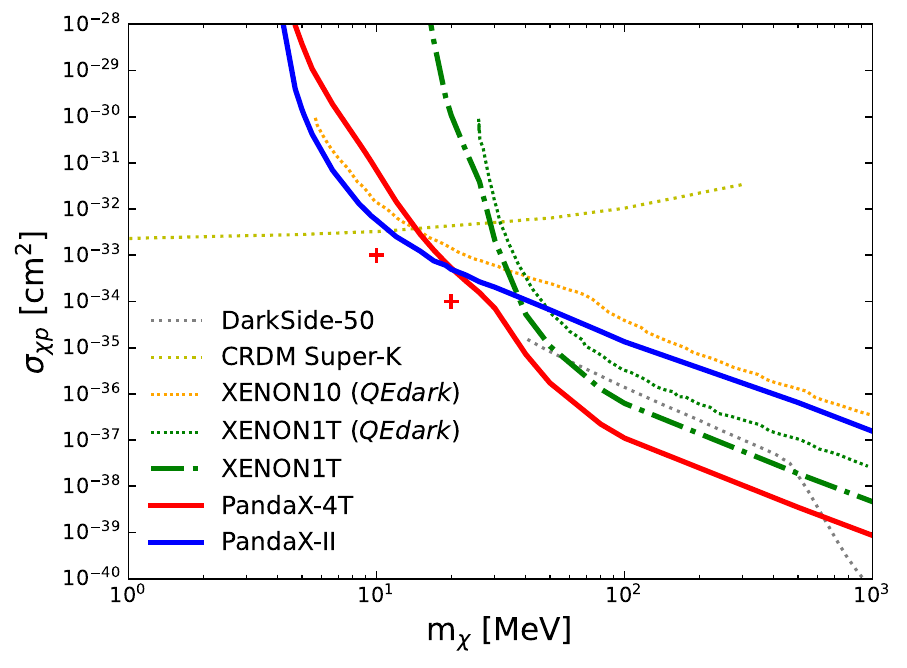}
    \includegraphics[width=\figsizeTwo]{\plotsdir/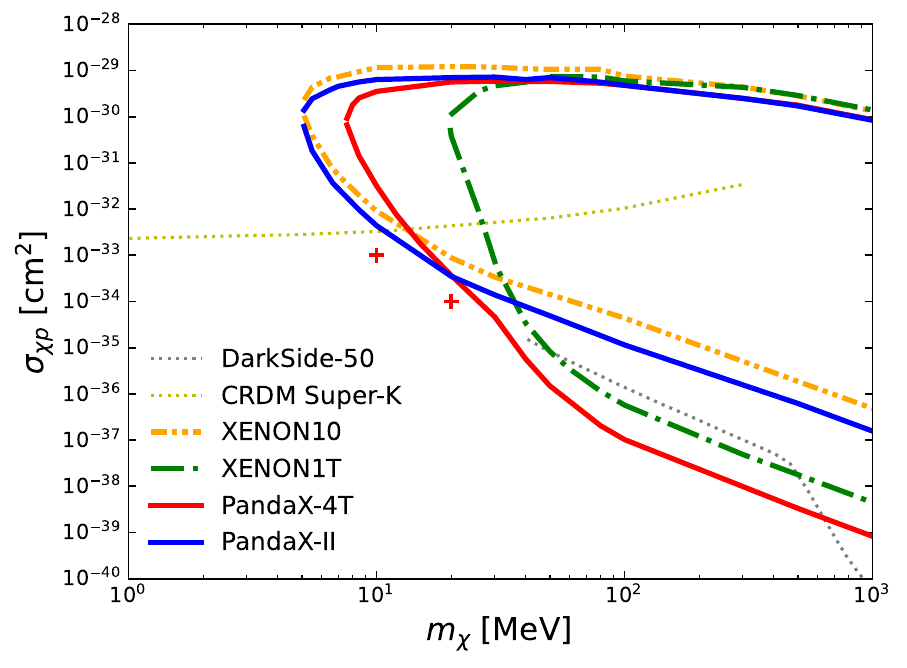}
    \caption{
        (Left)
        Constraints on SI DM-nucleon scattering cross sections at $90\,\%$ C.L.  using the Migdal effect from the S2-only data of  PandaX-II~(blue solid), PandaX-4T~(red solid),
        and XENON1T~(green dash-dotted) with the Earth shielding effect neglected.
The constraints from XENON10~(orange dotted) and XENON1T~(green dotted)~\cite{Essig:2019xkx} calculated using the \texttt{QEdark} code~\cite{QEdark} and the constraints from  DarkSide-50~\cite{DarkSide:2022dhx} on the Migdal effect of halo DM are shown.
The constraints from the data of Super-K on the cosmic-ray boosted DM are also shown for comparison~\cite{Super-Kamiokande:2022ncz}.   
The two red markers represent two benchmark parameter sets of DM particle masses and DM-nucleon scattering cross sections.
One is for that given in \eq{eq:typical_paras}, the other one is for  $m_\chi=20~\text{MeV}$, $\sigma_{\chi p}=10^{-34}~\text{cm}^2$.
(Right)
        The same as the left but with the Earth shielding effect included.
        The constraints from XENON1T and XENON10 with Earth shielding effect are calculated without using the \texttt{QEdark} code. 
}
    \label{fig_limits}
\end{figure}
In the calculation, we take $v_\oplus=232~{\text{km/s}}$ as a benchmark value~\cite{XENON:2018voc}  and neglect the orbital motion of the Earth around the Sun.
The results are shown in the left panel of~\fig{fig_limits}.  
We find that PandaX-II and PandaX-4T are currently placing the most stringent constraints on DM-nucleon scattering cross section for the DM particle mass below $\sim 600$~MeV.
While PandaX-II is leading the constraints in the DM mass range $\sim 11-20$~MeV,
the constraints from PandaX-4T are more stringent in the range $\sim 20-630$~MeV.
For higher DM particle mass from $630~\text{MeV}$ to $\sim 1~\text{GeV}$, the constraints on DM-nucleon scattering cross section are dominated by the results from Darkside-50~\cite{DarkSide:2022dhx}.
We also recalculate the constraints from the S2-only data of XENON1T~\cite{Aprile:2019xxb} using a more precise data analysis framework provided by the XENON1T collaboration~\cite{xenon1t:s2only_data_release}.
We obtain more stringent constraints than that previously obtained from the \texttt{QEdark} code~\cite{Essig:2019xkx}. However, the improved constraints of XENON1T are still about a factor of five weaker than that of PandaX-4T for DM particle mass from $\sim100~\text{MeV}$ to~$\sim1~\text{GeV}$.

In the next step, we include the effects of Earth shielding into the constraints. 
Since in the calculation of the constraints the time-integrated data with long exposure time are used, the time variation in the Earth shielding effect in a sidereal day can be neglected.
In the MC simulations of the DM propagation within the Earth,  the surface of the spheres where the underground laboratories are located is divided
linearly in $\cos\Theta$  from $\cos\Theta_{\min}$ to $\cos\Theta_{\max}$ into $20$ isodetection rings. We simulate $\mathcal{O}(10^5)$ DM trajectories passing through each isodetection ring.
In the right panel of~\fig{fig_limits}, we show the excluded regions on SI DM-nucleon scattering cross section derived from the S2-only data of PandaX-II and PandaX-4T.
We also recalculated the constraints from XENON1T and XENON10 using our MC simulations for Earth shielding effect. 
A significant difference from the case without Earth shielding effect is that
for SI DM-nucleon cross section larger than $\mathcal{O}(10^{-29})~\text{cm}^2$, 
the underground DM particles lose most of their kinetic energies due to the elastic scatterings within the Earth such that they cannot pass the detection thresholds of the experiments under consideration, which forms a detection blind zone for underground DM direct detections.
Furthermore,  it can be seen from \fig{fig_limits} that after including the Earth shielding  effect, 
the upper limits are also slightly stronger in the low mass region, which is related to the enhancement of the underground DM density due to the deflections of DM particles propagating within the Earth.
For instance, for PandaX-4T, the constraints on SI DM-nucleon cross section with and without the Earth shielding effect are $\sigma_{\chi p}\lesssim 3.2\times 10^{-32}~(3.7\times 10^{-34})~\text{cm}^2$ and $\sigma_{\chi p}\lesssim7.0\times 10^{-32}~(5.3\times 10^{-34})~\text{cm}^2$, respectively, for DM particle mass of $m_\chi=10~(20)$~MeV.
Similarly, for PandaX-II, the constraints on SI DM-nucleon cross section with and without the Earth shielding effect are $\sigma_{\chi p}\lesssim4.4\times 10^{-33}~(3.5\times 10^{-34})~\text{cm}^2$ and $\sigma_{\chi p}\lesssim5.8\times 10^{-33}~(5.1\times 10^{-34})~\text{cm}^2$, respectively, for DM mass of $10~(20)$~MeV.
Note that the benchmark values for the DM particle considered in \eq{eq:typical_paras} are still compatible with these updated constraints.

\section{Predictions for diurnal modulation amplitudes }\label{section_projection}
In this section, we make predictions for the diurnal modulation amplitudes of electron recoil signals induced by the Migdal effect for the experiments of PandaX-4T.
The final diurnal modulation amplitudes observed by the experiments strongly depend on the signal to background event ratio. 
Instead of predicting the time-variation curve of the electron events during a sidereal day, which requires very high statistics and should have similar time-variation patterns 
as that in \fig{fig_diurnal_effect_amp_PREM},
we consider the difference of the total event numbers $N(R_i)$ in two equal-length sidereal time intervals  $R_1=(\tau_1, \tau_1+\Delta\tau)$ and $R_2=(\tau_2, \tau_2+\Delta\tau)$ with different starting time $\tau_{1,2}$ as follows
\begin{align}
    A_R=\frac{N(R_1)-N(R_2)}{N(R_1)+N(R_2)} .
\end{align}
The observed events consist of signals and backgrounds, i.e., $N(R_i)=S(R_i)+B(R_i), \ (i=1,2)$.
For a given background event rate, it is necessary to optimize the sidereal time intervals to maximize the statistical significance of $A_R$.
For the background-dominated case, under the assumptions that the background event numbers $B(R_i)$ are constant in time and follow  Poisson distributions,
we find that a possible choice is  $\tau_1=16$~h sidereal~hour and $\tau_2=4$~h  sidereal~hour, and $\Delta\tau=9$ sidereal~hours. The two time intervals  $R_{1,2}$ are  illustrated in \fig{fig_diurnal_effect_amp_PREM}.

The Migdal effect is expected to be more significant towards lower PE regions. 
So far PandaX-4T has performed measurements in the region above 60~PE,
while the previous PandaX-II experiment has already measured the S2 signals down to 50~PE. 
PandaX-4T,  as a successor of PandaX-II,  should be capable to search for S2-only signals in lower PE regions as PandaX-II did in the near future. Thus, in this work, we focus on the predictions of diurnal modulation in the range of 50--55~PE for PandaX-4T.
Since the background event rate in the range 50-55~PE  is not yet estimated by PandaX-4T,
we use the posterior background event rate of  $b_{60}\approx1.3 \times10^{-2}~\text{/ton/day/PE}$ in the lowest S2 bin 60-70~PE as a reference value~\cite{PandaX:2022xqx}, and allow the background to vary by a few orders of magnitude.
We assume the same total detection efficiency of PMTs for S2-only signals ranging from 50-55~PE as that of 60~PE, since the total detection efficiency is almost a constant for S2-only signals ranging from 60-200~PE~\cite{PandaX:2022xqx}. 
 
\begin{figure}[t]
\includegraphics[width=\figsizeTwo]{\plotsdir/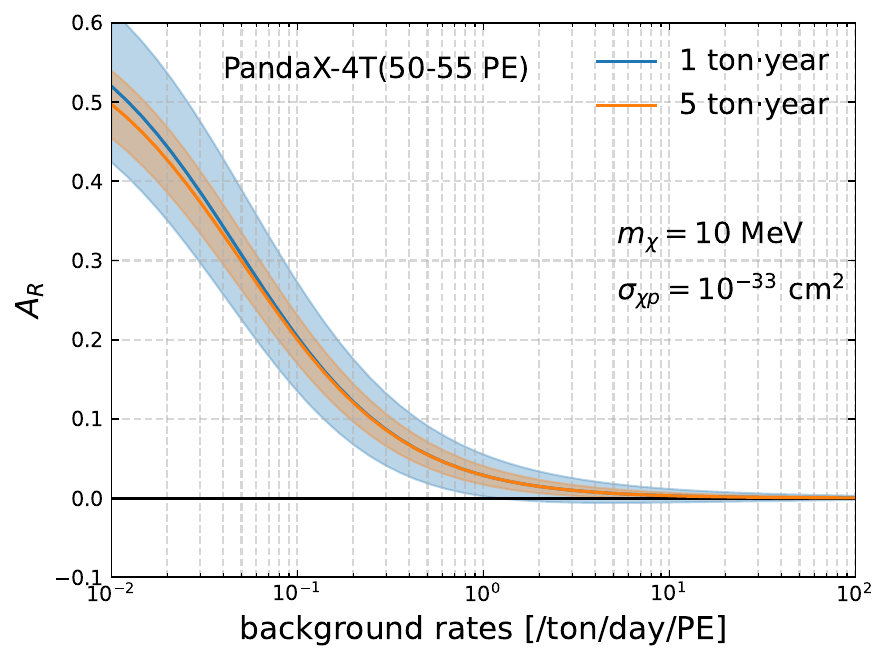}
    \includegraphics[width=\figsizeTwo]{\plotsdir/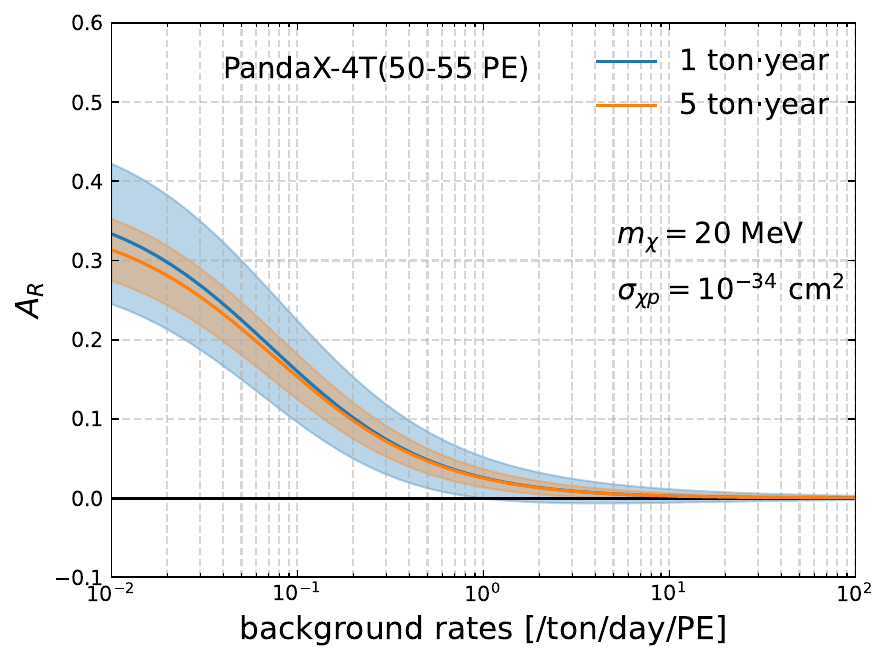}
    \caption{
(Left)
        The asymmetric parameter $A_R$ and its standard deviation as a function of background rates for the S2-only signals ranging from 50 to 55 PE of PandaX-4T with the benchmark parameters given by \eq{eq:typical_paras}.
The results with typical exposure of $1~\text{ton}\cdot\text{year}$~(118 sidereal days of PandaX-4T) and of $5~\text{ton}\cdot\text{year}$~(590 sidereal days of PandaX-4T) are shown in blue and orange, respectively.
(Right)
        The same as the left but for $m_\chi=20~\text{MeV}$, $\sigma_{\chi p}=10^{-34}~\text{MeV}$.
        }
    \label{fig_AR_significance}
\end{figure}

We calculate the S2-only signals induced by the Migdal effect with typical exposure time of $1~{\text{ton} \cdot \text{year}}$ (118 sidereal days) and 
$5~{\text{ton} \cdot \text{year}}$ (590 sidereal days)
based on the $3.1$~ton fiducial mass  of the PandaX-4T experiment~\cite{PandaX:2022xqx}.
The predicted events in the same sidereal hour on different sidereal days are summed together to increase the statistics.
In the calculation,  we take a  reference starting date of $t_0$ at 1st, Jan., 2024, 10:30:52 UT, which is the first zero hour of sidereal days in the year 2024 for CJPL where the PandaX-4T experiment is located.
In the left panel of \fig{fig_AR_significance},
 we show the predicted values of  $A_R$ and its standard deviation as a function of $b_{50}$, the background event rate at 50~PE, for the DM parameters in \eq{eq:typical_paras}.
It can be seen that the value of $A_R$  and also its statistic significance
 increase with decreasing background rate.
It is possible to observe a significant diurnal modulation of electron recoil signals even for $b_{50}$ is not too much higher than $b_{60}$.
For instance,  with $1~{\text{ton} \cdot \text{year}}$  of PandaX-4T data, we find that 
 the asymmetry can reach
 \begin{align*}
    A_R=(2.11 \pm 0.70) \times 10^{-1} \quad \text{for} \quad
    b_{50}=9.5 \times 10^{-2}~\text{/ton/day/PE},
 \end{align*}
which suggests that if the background at 50~PE is not too high, namely, $b_{50}/b_{60}\lesssim 7.3$, 
 the diurnal modulation can be around $3\sigma$ above zero. 
With increasing data taking of $5~{\text{ton} \cdot \text{year}}$, the predicted asymmetry can reach 
$A_R=(4.48 \pm 1.47) \times 10^{-2}$ which is around $3\sigma$ above zero for a higher  background rate $b_{50}= 6.2 \times 10^{-1}~\text{/ton/day/PE}$.
For the background-dominated case that we are concerned about, the central value of $A_R$ is insensitive to the starting point $t_0$.
However, at very  low backgrounds, the predicted central value of $A_R$ with different exposures
are slightly different, which is related to the dependence on the starting date $t_0$.
These differences result from the slow variation of $\boldsymbol{v}_\oplus$ during a longer exposure time as indicated in \fig{fig_AR_significance}.

In the right panel of \fig{fig_AR_significance}, we also show $A_R$ and its standard deviation for a different DM mass $m_\chi=20~\text{MeV}$ and a smaller DM-nucleon cross section $\sigma_{\chi p}=10^{-34}~\text{cm}^2$ which is also consistent with the updated constraints.
Due to the smaller DM-nucleon cross section than that given by \eq{eq:typical_paras}, the diurnal modulation under this parameter choice is weaker.
With $1~{\text{ton} \cdot \text{year}}$ of PandaX-4T data,
the predicted asymmetry can reach
\begin{align*}
    A_R=(2.29 \pm 0.76) \times 10^{-1} \quad \text{for} \quad
    b_{50}=4.8 \times 10^{-2}~\text{/ton/day/PE},
    \end{align*}
which is around $3\sigma$ above zero for $b_{50}/b_{60}\lesssim 3.7$.
With increasing data taking of $5~{\text{ton} \cdot \text{year}}$, 
we find that the asymmetry can reach
$A_R=(5.06 \pm 1.66) \times 10^{-2}$
which is around $3\sigma$ above zero for a higher background rate lower than $b_{50}=4.6 \times 10^{-1}~\text{/ton/day/PE}$.

\section{Conclusion}\label{section_conclusion}
In summary, we have discussed a novel type of diurnal modulation effect, diurnal modulation in electron recoil signals induced by DM-nucleon scattering via the Migdal effect.  To our knowledge,  it is the only possible mechanism to give rise to significant diurnal modulation in electron events given the current stringent experimental constraints.
We have updated the constraints on the Midgal effects using the S2-only data of  PandaX-II and PandaX-4T with improved Monte-Carlo simulations of the Earth shielding effect, which sets so far the most stringent constraints on  DM-nucleon scattering cross section via the Migdal effect for DM particle mass below $\sim 1$~GeV.
Based on the news constraints, we predict that the Migdal effect induced diurnal modulation of electron recoils can still be significant in the low energy region. 
For instance, for $1~{\text{ton} \cdot \text{year}}$  of PandaX-4T data, we have found that with at a background rate $\lesssim 9.5 \times 10^{-2}~\text{/ton/day/PE}$ in the S2 signal bin at 50-55~PE, the predicted asymmetry is $A_R\approx (2.11 \pm 0.70) \times 10^{-1}$,  which is around $3\sigma$ above zero. 
These predictions in the low S2 bins can be tested by the  PandaX-4T experiment in the near future.
Extending our analysis to the other xenon-based experiments such as XENONnT and LZ is straightforward.

  \begin{acknowledgments}
    We are grateful to Jianglai Liu, Yong Yang, and Ning Zhou for their helpful discussions on the PandaX data analysis.
This work is supported in part by
the National Key R\&D Program of China No.~2017YFA0402204,
the National Natural Science Foundation of China (NSFC)
    No.~11825506,  No.~11821505,  No.~12047503,  and
    No.~12247148. \end{acknowledgments}

\appendix
\section{Sidereal time, celestial coordinates and the velocity of the Earth}\label{appendix_sidereal}
This appendix provides a brief review of the transformation from the Universal Time~(UT) to the  local sidereal time,
the definitions and transformations of celestial coordinates,
and the variation of the Earth velocity in the galactic rest frame, $\boldsymbol{v}_\oplus$, over the course of a year, based on \REF{Emken:2017qmp, McCabe:2013kea}.

\subsection{Sidereal time}
The length of a sidereal day is $T_\tau=86164.1$~s~\cite{ParticleDataGroup:2022pth}, which is defined as the period of the self-rotation of the Earth.
We introduce a sidereal second $\text{s}_\tau=T_\tau/86400$ and a sidereal hour $\text{h}_\tau=T_\tau/24$, which is different from a mean solar hour $\text{h}=3600~\text{s}$.
The zero hour of a sidereal day is the time point when the vernal equinox comes across the local meridian.
For given Universal Time $t$ and location, the local sidereal time $\tau$ can be uniquely determined.

For a time point $t=D,\,M,\,Y,\,$~$h$:$m$:$s$~UT, the fractional number of days $n_\text{J2000.0}$ relative to the reference time $\text{J2000.0}$,
which is 1st,~Jan.,~2000,~12:0:0 of territorial time~(TT), 
is given by~\cite{McCabe:2013kea}
\begin{align}
    \begin{split}
        n_\text{J2000.0}=&\text{floor}\lra{365.25\tilde{Y}}+\text{floor}\lra{30.61\lra{\tilde{M}+1}}+D\\
        &+\frac{h}{24}+\frac{m}{24\times 60}+\frac{s}{24\times 60\times 60}-730563.5,
    \end{split}
    \end{align}
where
the function $\text{floor}(x)$ returns the largest integer less than or equal to $x$,
$\tilde{Y}=Y-1$ and $\tilde{M}=M+12$ for January and February, respectively,
while $\tilde{Y}=Y$ and $\tilde{M}=M$ for the other months,
and $h$, $m$, $s$, $D$, $M$, and $Y$ are the hour, minute, second, day, month, and year of the Universal Time, respectively.
The Greenwich apparent sidereal time~(GAST) for a given $n_\text{J2000.0}$ 
is approximately given by~\cite{Emken:2017qmp}
\begin{align}
\begin{split}
    \text{GAST}
    \simeq
    &
    \left(
    86400
    \lra{
        0.78 
        +
        \text{Mod}\lra{n_{\text{J2000.0}},\,1}
        + 0.0027n_{\text{J2000.0}}
    }
    +
    9.7 \times 10 ^ {-4}
    \right.
    \\
    &
    \left.
    +
    307.48 T_\text{J2000.0}
    + 
    0.093 T_\text{J2000.0}^2
+
    \lra{ -1.15 \sin \Omega - 0.086 \cos 2L}
    \cos \epsilon_A
    \right.
    \\
    &
    \left.
    + 
    1.76\times 10^{-4} \sin\Omega + 4\times 10^{-6} \sin 2\Omega
    \right)
    \text{s}_\tau
    ,
\end{split}
\end{align}
where 
the function $\text{Mod}\lra{x,\,y}$ gives the remainder after the division of $x$ over $y$, 
$T_\text{J2000.0}={n_\text{J2000.0}}/{36525}$,
$\Omega \simeq 125.04^\circ - 0.053^\circ n_\text{J2000.0}$,
$L\simeq 280.47^\circ+0.99^\circ n_\text{J2000.0}$,
and $\epsilon_A \simeq 23.44^\circ - 0.013 ^\circ T_\text{J2000.0}$.
Then, the sidereal time at Universal Time $t$ for a laboratory with longitude $\lambda_\text{lab}$ is given by~\cite{Emken:2017qmp}
\begin{align}
    \tau
    =
    \text{Mod}
    \lra{
        \text{GAST}
        \lra{t}
        +
        \frac{\lambda_\text{lab}}{2\pi}
        T_\tau,
        \,
        T_\tau
    },
    \label{eq:LAST}
\end{align}
where $\lambda_\text{lab}$ is positive for eastern longitudes but negative for western longitudes
and $\tau \in (0,\,24)~\text{h}_\tau$.

\begin{figure}[t]
    \centering
        \includegraphics[width=\figsizeOne]{\plotsdir/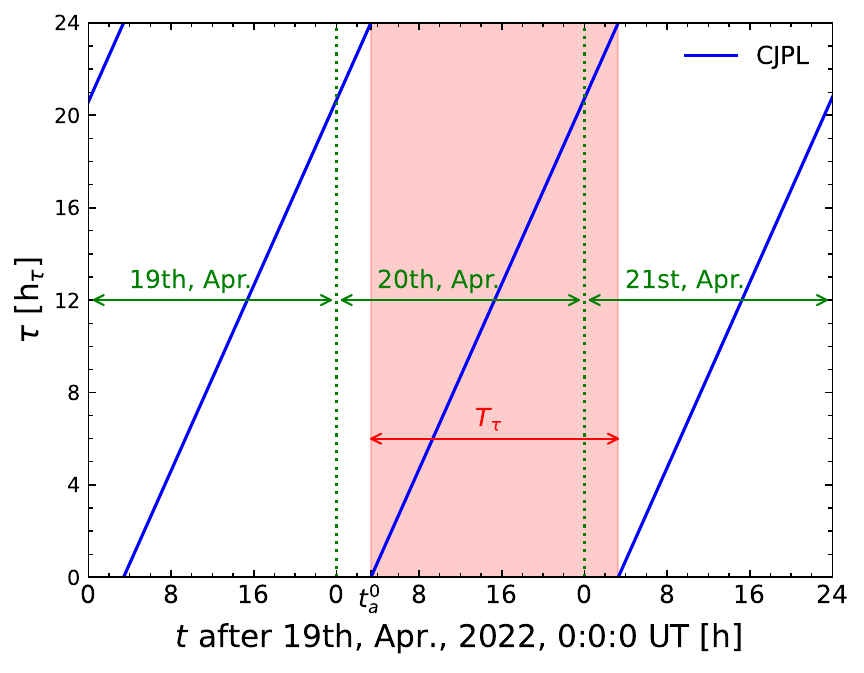}
    \caption{
    Sidereal time $\tau$ at CJPL for Universal Time $t$ after $19\text{th}$, Apr., 2022,~0:0:0~UT.
A sidereal day on the typical date $t_a$ starting from $t_{a}^0=20\text{th}$,~Apr.,~2022,~3:20:24~UT~(red shaded) is shown.
    }\label{fig_t_ts}
\end{figure}

In \fig{fig_t_ts}, we present the corresponding sidereal time $\tau$ at CJPL since $19\text{th}$,~Apr.,~2022,~0:0:0~UT as an example.
We highlight a typical date $t_{a}=20\text{th}$,~Apr.,~2022~UT, which corresponds to the maximum declination $\delta_\text{halo}=-41^\circ$ of the velocity of the DM halo in the Earth's rest frame in the year 2022.
We show the sidereal day on $t_a$ starting from $t_{a}^0=20\text{th}$,~Apr.,~2022,~3:20:24~UT in \fig{fig_t_ts}, which we use for diurnal modulation analysis in this work.
Due to the difference in longitudes between CJPL and LNGS, there is a time shift of approximately 5.9~h in the relation between $t$ and $\tau$ at LNGS compared to CJPL. 

\subsection{Celestial coordinates and transformations}
We briefly review the transformations between celestial coordinates including the heliocentric ecliptic, equatorial, and galactic coordinates in this section based on~\REF{Emken:2017qmp,McCabe:2013kea}.
The origins of the heliocentric ecliptic, equatorial, and galactic coordinate systems are all referenced to the center of the Earth.
The $z$-direction of the equatorial and galactic coordinates are pointed to the North Celestial Pole~(NCP) and the North Galactic Pole~(NGP), respectively.
The $x$-direction of the equatorial and heliocentric ecliptic coordinates are both pointed to the vernal equinox.
As for the galactic coordinates, the $x$-direction is pointed to the galactic center.

The transformation from the heliocentric ecliptic coordinates into the equatorial coordinates at $T_\text{J2000.0}$ is given by $\boldsymbol{x}^\text{ hel, ecl}=-\mathcal{R}\boldsymbol{x}^\text{gal}$, where the matrix $R$ is given by
    \begin{align}
        \mathcal{R}=
        \begin{pmatrix}
            1 & 0 & 0\\
            0 & \cos\epsilon & -\sin\epsilon\\
            0 & \sin\epsilon & \cos\epsilon
        \end{pmatrix}
        ,
    \end{align}
and $\epsilon=23.44^\circ-0.013^\circ T_\text{J2000.0}$ is the obliquity of the ecliptic~\cite{Emken:2017qmp}.
The transformation from equatorial coordinates into galactic coordinates at $\text{J2000.0}$ is given by $\boldsymbol{x}^\text{gal}=\mathcal{M}\boldsymbol{x}^\text{equ}\lra{\text{J2000.0}}$, where $\mathcal{M}$ is a $3\times 3$ matrix. The elements of matrix $\mathcal{M}$ are given by
\begin{align}
    \begin{split}
        \mathcal{M}_{11}&=-\sin l_\text{NCP}\sin \alpha_\text{NGP}-\cos l_\text{NCP}\cos\alpha_\text{NGP}\sin\delta_\text{NGP},\\
        \mathcal{M}_{12}&=\sin l_\text{NCP}\cos \alpha_\text{NGP}-\cos l_\text{NCP}\sin\alpha_\text{NGP}\sin\delta_\text{NGP},\\
        \mathcal{M}_{13}&=\cos l_\text{NCP}\cos \alpha_\text{NGP},\\
        \mathcal{M}_{21}&=\cos l_\text{NCP}\sin \alpha_\text{NGP}-\sin l_\text{NCP}\cos\alpha_\text{NGP}\sin\delta_\text{NGP},\\
        \mathcal{M}_{22}&=-\cos l_\text{NCP}\cos \alpha_\text{NGP}-\sin l_\text{NCP}\sin\alpha_\text{NGP}\sin\delta_\text{NGP},\\
        \mathcal{M}_{23}&=\sin l_\text{NCP}\cos \delta_\text{NGP},\\
        \mathcal{M}_{31}&=\cos\alpha_\text{NGP}\cos\delta_\text{NGP},\\
        \mathcal{M}_{32}&=\sin\alpha_\text{NGP}\cos\delta_\text{NGP},\\
        \mathcal{M}_{33}&=\sin\delta_\text{NGP},
    \end{split}
    \end{align}
where $\delta_\text{NGP}=27.1^\circ$ and $\alpha_\text{NGP}=192.9^\circ$ are the declination and right ascension of the North Galactic Pole at $\text{J2000.0}$, respectively,
and $l_\text{NCP}=122.9^\circ$ is the galactic longitude of the North Celestial Pole at $\text{J2000.0}$~\cite{Emken:2017qmp}.
The transformation of a vector at $\text{J2000.0}$ into that at $T_\text{J2000.0}$ in equatorial coordinates is given by $\boldsymbol{x}^\text{equ}\lra{T_\text{J2000.0}}=\mathcal{P}\boldsymbol{x}^\text{equ}\lra{\text{J2000.0}}$,
where the elements of matrix $\mathcal{P}$ are given by
\begin{align}
    \begin{split}
        \mathcal{P}_{11}&=\cos\zeta_A\cos\theta_A\cos z_A-\sin\zeta_A\sin z_A,\\
        \mathcal{P}_{12}&=-\sin\zeta_A\cos\theta_A\cos z_A-\cos\zeta_A\sin z_A,\\
        \mathcal{P}_{13}&=-\sin\theta_A\cos z_A,\\
        \mathcal{P}_{21}&=\cos\zeta_A\cos\theta_A\sin z_A+\sin\zeta_A\cos z_A,\\
        \mathcal{P}_{22}&=-\sin\zeta_A\cos\theta_A\sin z_A+\cos\zeta_A\cos z_A,\\
        \mathcal{P}_{23}&=-\sin\theta_A\sin z_A,\\
        \mathcal{P}_{31}&=\cos\zeta_A\sin\theta_A,\\
        \mathcal{P}_{32}&=-\sin\zeta_A\sin\theta_A,\\
        \mathcal{P}_{33}&=\cos\theta_A,
    \end{split}
    \end{align}
with $\zeta_A=2306.08^{''}T_\text{J2000.0}+0.30^{''}T_\text{J2000.0}^2$,
$\theta_A=2306.08^{''}T_\text{J2000.0}+1.09^{''}T_\text{J2000.0}^2$,
and $z_A=2004.19^{''}T_\text{J2000.0}+0.43^{''}T_\text{J2000.0}^2$~\cite{Emken:2017qmp}.
Then, the transformation from the heliocentric ecliptic coordinates into the galactic coordinates and that from the galactic coordinates into the equatorial coordinates at $T_\text{J2000.0}$ can be given by $\boldsymbol{x}^\text{gal}=-\mathcal{M}\mathcal{P}^{-1}\mathcal{R}\boldsymbol{x}^\text{ hel, ecl}$ and $\boldsymbol{x}^\text{equ}\lra{T_\text{J2000.0}}=\mathcal{P}\mathcal{M}^{-1}\boldsymbol{x}^\text{gal}$, respectively.

\subsection{The Earth velocity in galactic rest frame}\label{appendix_vE}
The velocity of the Earth in the galactic rest frame is given by~\cite{McCabe:2013kea}
    \begin{align}
        \boldsymbol{v}_\oplus=\boldsymbol{v}_r+\boldsymbol{v}_s+\boldsymbol{v}_e\lra{t},
    \end{align}
where $\boldsymbol{v}_r=\lra{0,~220,~0}^{\text{T}}~\text{km/s}$ is the galactic rotation velocity with $\text{T}$ stands for the transpose of a matrix, 
$\boldsymbol{v}_s=\lra{11.1,~12.2,~7.3}^{\text{T}}~\text{km/s}$ is the velocity of the motion of the sun relative to the nearby stars,
and $\boldsymbol{v}_e\lra{t}$ is the orbital velocity of the Earth relative to the sun in galactic coordinates given by~\cite{McCabe:2013kea}
\begin{align}
        \boldsymbol{v}_e\lra{t}=-\lre{v_e}\lra{
            \lra{\sin L'+e\sin\lra{2L'-\overline{\omega}}}\boldsymbol{e}_x
            -\lra{\cos L'+e\cos\lra{2L'-\overline{\omega}}}\boldsymbol{e}_y
            },
    \end{align}
with $\lre{v_e}=29.79~\text{km/s}$,
$e=0.017$,
$L'=280.46^\circ+0.99^\circ n_\text{J2000.0}$,
$\overline{\omega}=282.93^\circ+0.000047^\circ n_\text{J2000.0}$,
$\boldsymbol{e}_x=-\mathcal{M}\mathcal{P}^{-1}\mathcal{R}\lra{1,0,0}^\text{T}$,
and $\boldsymbol{e}_y=-\mathcal{M}\mathcal{P}^{-1}\mathcal{R}\lra{0,1,0}^\text{T}$.
Then, we transform $\boldsymbol{v}_\oplus$ from the galactic coordinates into the equatorial coordinates.

\begin{figure}[t]
    \centering
    \subfigure[~$\delta_{\oplus}$]{
        \includegraphics[width=\figsizeThree]{\plotsdir/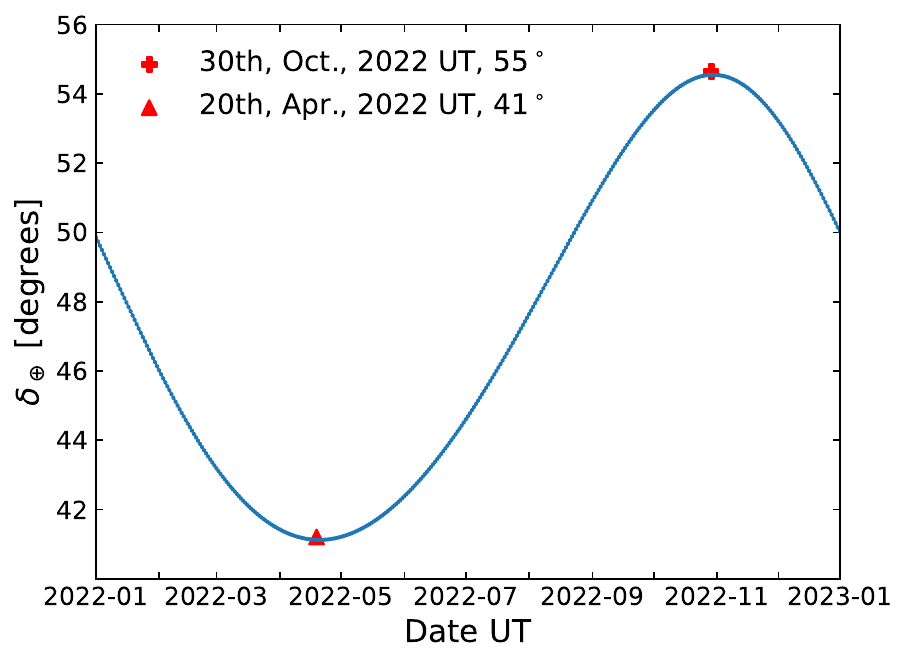}
    }
    \subfigure[~$\alpha_{\oplus}$]{
        \includegraphics[width=\figsizeThree]{\plotsdir/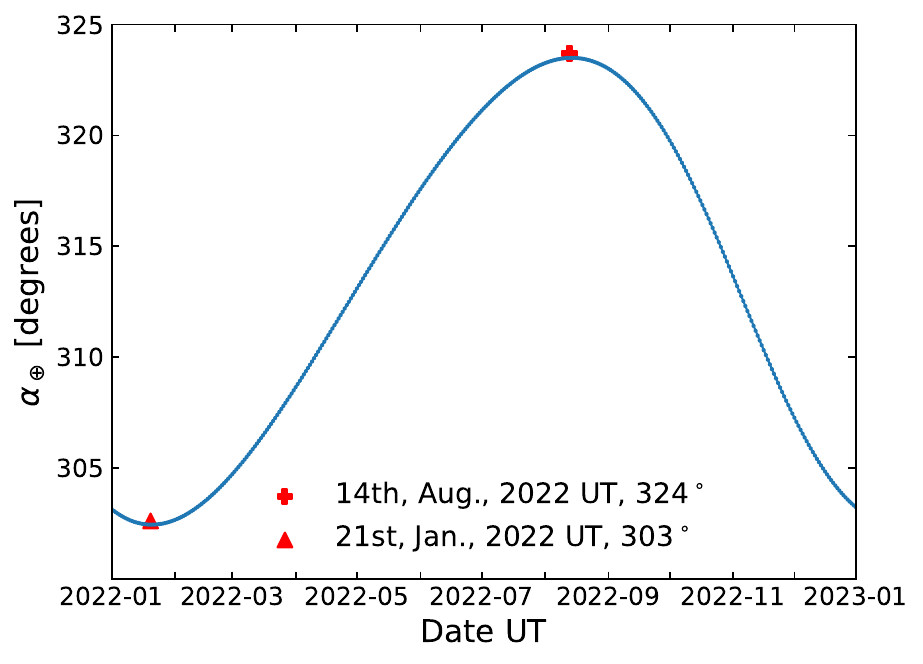}
    }
    \subfigure[~$v_{\oplus}$]{
        \includegraphics[width=\figsizeThree]{\plotsdir/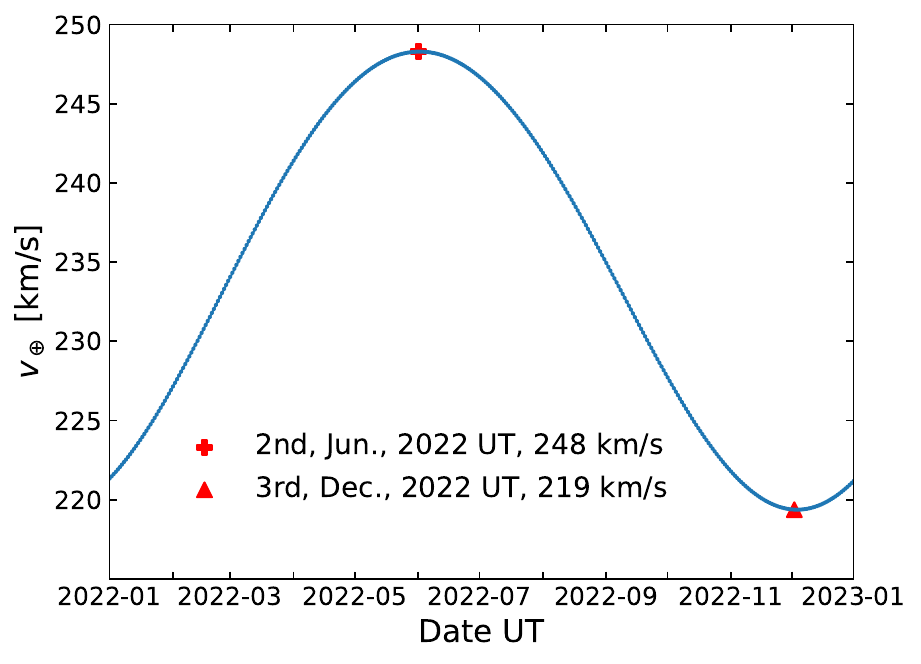}
}
    \caption{Annual modulation of $\delta_\oplus$, $\alpha_\oplus$, and $v_\oplus$ during the year 2022.
The red markers stand for the maximum or minimum value of $\delta_\oplus$, $\alpha_\oplus$, and $v_\oplus$ during the year 2022.
}\label{fig_vE_deltaE_alphaE_annual}
\end{figure}

In \fig{fig_vE_deltaE_alphaE_annual}, we show the variation of $\boldsymbol{v}_{\oplus}$ in the equatorial coordinate during the year 2022 as an example.
With the parameters taken from~\cite{Emken:2017qmp,timon_emken_2020_3726878},
the declination $\delta_\oplus$, right ascension $\alpha_\oplus$, and the magnitude $v_{\oplus}$ of $\boldsymbol{v}_{\oplus}$ varies from $41^\circ$ to $ 55^\circ$, $303^\circ$ to $ 324^\circ$, and $219~\text{km/s}$ to $248~\text{km/s}$, respectively, during the year 2022.
The relation between the equatorial coordinates of the velocity of DM halo $\boldsymbol{v}_{\text{halo}}$ in the rest frame of the Earth and $\boldsymbol{v}_{\oplus}$ are given by $\delta_{\text{halo}}=-\delta_\oplus$ and $\alpha_{\text{halo}}=\alpha_\oplus-\pi$.

\section{Details on MC simulations}\label{appendix_mc_details}
\subsection{Initial conditions}\label{appendix_mc_details_ini}
\begin{figure}[t]
        \centering
        \includegraphics[width=5 cm]{\plotsdir/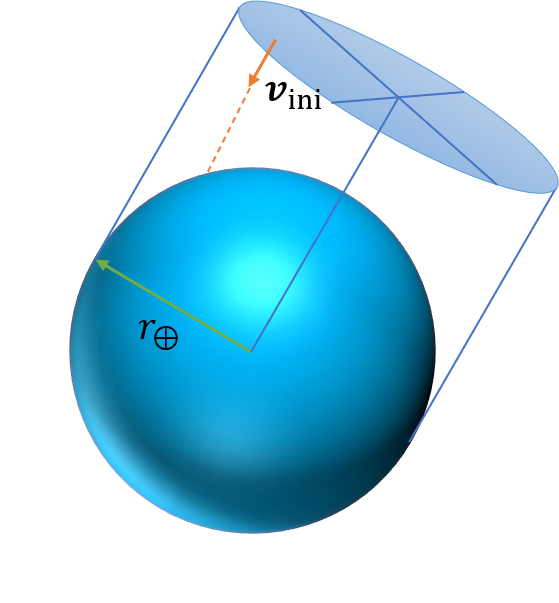}
        \caption{
            Sketch for the initial position of a DM particle with the initial velocity $\boldsymbol{v}_\text{ini}$ in the rest frame of the Earth.
        }
        \label{fig_injection}
    \end{figure}

We sample the initial velocities of halo DM particles isotropically in the rest frame of the galaxy according to the velocity distribution $f_\text{halo}\lra{\boldsymbol{v}}$, and then subtract the velocity of the Earth in the galactic rest frame to obtain the initial DM velocity, $\boldsymbol{v}_\text{ini}$, in the rest frame of the Earth.
We restrict DM speeds of the samples in the rest frame of the Earth range from $v_\text{th}$ to $v_{\max}$, where $v_{\max}=v_\text{esc} + v_\oplus$ with $v_\text{th}$ taken as $30~\text{km/s}$, below which the sub-GeV DM can not exceed the detection threshold of current DD experiments.
We sample the speeds of DM particles according to speed distribution instead of flux.
Therefore, each particle with initial speed $v_\text{ini}$ should be assigned a weight of $v_\text{ini} N_2 / N_1$,
where $N_1=\int_{v_\text{th}}^{v_{\max}}vf\lra{v}dv$, $N_2=\int_{v_\text{th}}^{v_{\max}}f\lra{v}dv$, and $f\lra{v}=\int v^2f_\text{halo}\lra{\boldsymbol{v}+\boldsymbol{v}_\oplus}d\Omega$.
For each DM particle, the initial position is sampled uniformly on a circular disc of radius $r_\oplus$ perpendicular to $\boldsymbol{v}_\text{ini}$ as illustrated in \fig{fig_injection}. 

\subsection{Underground DM speed distributions on different isodetection rings}\label{appendix_mc_details_dis}
We record the events when DM particles come across the isodetection rings at a given depth with speeds above $v_\text{th}$ in the rest frame of the Earth.
The events on each isodetection ring that we record in a simulation are equivalent to that recorded in an effective time interval $\Delta t$ in the case of continuous injection of DM~\cite{Xia:2021vbz}.
The effective time is determined by the number of injected DM particles $N_\text{sim}$ and the initial DM flux $N_1$ as $\Delta t=N_\text{sim}/\lra{\pi r_\oplus^2N_1}$.
Then, the underground DM flux on an isodetection ring at a depth $d$ can be expressed as
    \begin{align}
        \int v^3f\lra{\boldsymbol{v},\,d,\,\Theta}\lrd{\cos\theta} d\Omega=\frac{\Delta N}{\Delta v \Delta t \Delta S},\label{eq_flux}
    \end{align}
where $f\lra{\boldsymbol{v},\,d,\,\Theta}$ is the underground DM velocity distribution function on the isodetection ring with isodetection angle $\Theta$ at the depth $d$ normalized to the DM number density,
$\Delta N$ is the number of events crossing the isodetection ring with an area $\Delta S=2\pi \lra{r_\oplus-d}^2\lra{\cos\Theta-\cos\lra{\Theta+\Delta\Theta}}$ in a time interval $\Delta t$ and speed interval $[v,~v+\Delta v]$,
and $\theta$ is the angle between the DM velocity $\boldsymbol{v}$ and the normal direction of the isodetection ring at the crossing point.

To reconstruct the underground speed distribution,
each event $j$ should have a weight of $1/v_j\lrd{\cos\theta_j}$.
Combining the weight from the initial condition,
each event acquires a weight of $w_j={N_2v_\text{ini}}/N_1v_j\lrd{\cos\theta_j}$ in the reconstruction of the probability distribution function $p\lra{v,\,d,\,\Theta}$, which is normalized to 1, for DM speeds above $v_\text{th}$ on an isodetection ring at a depth $d$.
Finally, the speed distribution $f\lra{v,\,d,\,\Theta}$ can be expressed as
\begin{align}
        f\lra{v,\,d,\,\Theta}=
        \frac{p\lra{v,\,d,\,\Theta}}{\Delta t \Delta S}
        \sum_j^{}w_j.
        \label{eq_iso_distribution}
    \end{align}
It should be noted that the final expression of \eq{eq_iso_distribution} does not contain the normalization factor $N_1$,
since $N_1$ in $\Delta t$ and $w_j$ cancels out.

\section{Details on data analysis procedures of PandaX-II}\label{appendix_PandaX-II}
In this appendix, we briefly review the data analysis procedures of PandaX-II for electron recoils with MC simulation based on~\REF{Cheng:2021fqb, PandaX-II:2020oim, PandaX-II:2021jmq, Essig:2017kqs} and cross-check constraint on DM-electron cross section with that of PandaX-II~\cite{Cheng:2021fqb}.

For DM-electron scattering, following \REF{Essig:2017kqs, PandaX-II:2021jmq}, a primary ionized electron with kinetic energy $T_e$ can generate electron quanta in liquid xenon, which is described by the charge yield $Q_y$ for electron recoils.
The primary ionized electron itself can also contribute as a single electron quantum with probability $1-r$,
where $r$ is the mean recombination probability and almost vanishes at low kinetic energies for electron recoils~\cite{Sorensen:2011bd, Essig:2012yx, Essig:2017kqs}.
The number of electron quanta $n_e$ is generated by a normal distribution
$
    \text{norm}\lra{n_e^0,\,f_l \cdot n_e^0},
$
where $n_e^0=1-r+Q_y (T_e+n_2W)$ is the mean value of $n_e$,
$W=13.7~\text{eV}$~\cite{Cheng:2021fqb, PandaX-II:2020oim, PandaX-II:2021jmq} is the average energy required to create a single quantum in liquid xenon,
$n_2$ is the minimal additional quanta contributed by the de-excitation of the target atom in liquid xenon taken from~\cite{Essig:2017kqs} as listed in \tab{tab_add_quanta_Xenon},
and $f_l$ describes the variation of $n_e$~\cite{PandaX-II:2021jmq}.
\begin{table}\centering
    \begin{tabular}{@{}cccccc}
        \toprule
        Shell & $5p^6$ & $5s^2$ & $4d^{10}$ & $4p^6$ & $4s^2$ \\
        \colrule
        Binding energy [eV] & 12.4 & 25.7 & 75.6 & 163.5 & 213.8\\
        \colrule
        Minimal photon energy [eV] & 0 & 13.3 & 63.2 & 87.9 & 50.3\\
        \colrule
        Minimal additional quanta &  0 & 0 & 4 & 6 & 3 \\
        \botrule
    \end{tabular}
    \caption{Minimal additional quanta contributed by the de-excitation of different shells of xenon atoms from~\cite{Essig:2017kqs}.}\label{tab_add_quanta_Xenon}
\end{table}

The electron quanta can drift into the surface between liquid and gaseous xenon under the drift electric field with survival probability $s=\exp\lra{-t_\text{dri}/\tau_e}$~\cite{PandaX-II:2021jmq},
where $t_\text{dri}$ is the drift time of an electron
and $\tau_e=706~\mu\text{s}$~\cite{PandaX-II:2021jmq}.
The maximal values of $t_\text{dri}$ and the corresponding minimal $s$ for each scientific run of PandaX-II are listed in \tab{tab_drift_time}~\cite{PandaX-II:2021jmq}.
\begin{table}\centering
    \begin{tabular}{@{}cccc}
        \toprule
        Data set & run9 & run10 & run11 \\
        \colrule
        $t_\text{dri}^{\max}$ [$\mu s$] & 350 & 360 & 360\\
        \colrule
        $s_{\min}$ & 0.61 & 0.60 & 0.60 \\
        \botrule
    \end{tabular}
    \caption{Maximal drift time $t_\text{dri}^{\max}$ and minimal survival probability $s_{\min}$ from~\cite{PandaX-II:2021jmq}.}\label{tab_drift_time}
\end{table}
According to \tab{tab_drift_time}, the mean value of $s$ should lie somewhere between $0.6$ and $1.0$.
The number of the electron quanta $N^\prime_e$ that drift into the surface between liquid and gaseous xenon is generated by a binomial distribution $\text{binom}(n_e,\,s)$.

Then, the electron quanta that have successfully drifted to the liquid-gas surface can be extracted into xenon gas by the extraction electric field and induce scintillation which is known as the S2 signals.
The number of the electron quanta $N_e^{\prime\prime}$ extracted into xenon gas is generated by a binomial distribution $\text{binom}(N_e^\prime,\,\text{EEE})$,
where $\text{EEE}$ is the electron extraction efficiency~\cite{Cheng:2021fqb}.
The raw S2 signals induced in gaseous xenon ${\text{S2}_\text{raw}}$ are generated by a normal distribution
$
    \text{norm}(N_e^{\prime\prime}\cdot\text{SEG},\,\sigma_\text{SE}\sqrt{N_e^{\prime\prime}}),
    $
where $\text{SEG}$ and $\sigma_\text{SE}$ are the single-electron gain and its resolution~\cite{Cheng:2021fqb}, respectively.
The $\text{S2}_\text{raw}$ signals generated in xenon gas can be received by the detector as the $\text{S2}$ signals with a probability which is taken as the efficiency from~\cite{Cheng:2021fqb}.
In addition, due to the nonlinear effect of baseline suppression firmware~\cite{PandaX-II:2020oim, PandaX-II:2021jmq}, the final S2 signals are given by ${\text{S2}}\cdot f_2\lra{\text{S2}}$,
where the value of S2-dependent function $f_2$ is taken from~\cite{PandaX-II:2020oim}.

The value of $f_l$, which describes the variation of $n_e$, for $T_e\lesssim 1~{\text{keV}}$
and the mean value of the survival probability $s$ were not explicitly published by the PandaX-II collaboration.
For simplicity, we treat both $f_l$ and $s$ as constants and tune their values by cross-checking our constraints on the DM-electron scattering cross section through contact interactions with the results reported by PandaX-II~\cite{Cheng:2021fqb}.

We adopt the event rate of DM scattering off bound electrons described in~\REF{Essig:2017kqs, Essig:2019xkx}.
We take $v_\oplus=232~\text{km/s}$~\cite{XENON:2018voc}.
We extract the numerical results of the ionization factor of xenon atoms from the \texttt{QEdark}~\cite{QEdark} code as the input data to calculate the event rates of DM-electron scatterings in this work.
We set the constraints with the Binned Poisson method~\cite{Savage:2008er, Green:2001xy} at $90\,\%$ C.L. using the S2-only data of PandaX-II in 5 bins ranging from $50$ to $75$~PE~\cite{Cheng:2021fqb}.
We neglect the background rates due to the lack of complete estimation of the background available from PandaX-II.

\begin{figure}[t]
        \centering
        \includegraphics[width=\figsizeOne]{\plotsdir/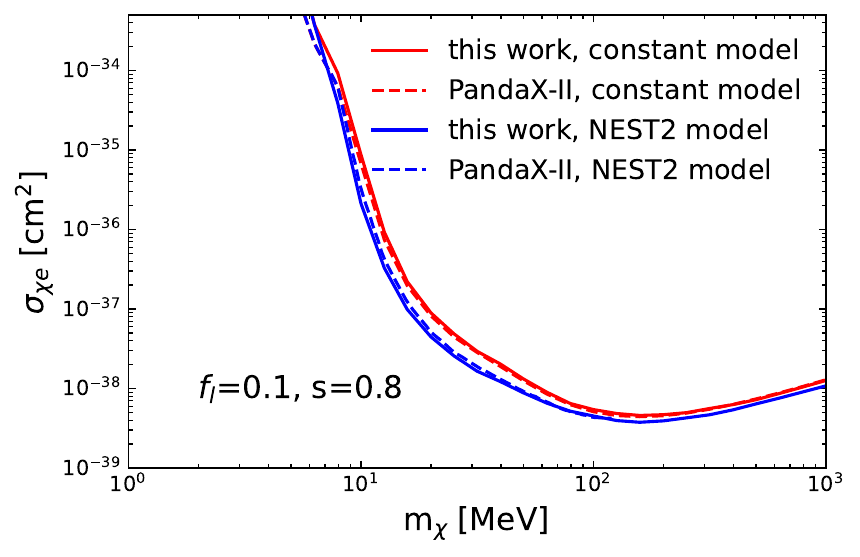}
        \caption{
            Constraints on DM-electron scattering cross sections at $90\,\%$ C.L. derived from S2-only data of PandaX-II using the constant charge yield model~(red solid) and the NEST2 charge yield model~(blue solid) with $f_l = 0.1$ and $s=0.8$.
            The results with the constant~(red dashed) and the NEST2 charge yield model~(blue dashed) from PandaX-II~\cite{Cheng:2021fqb}~are shown for comparison.
        }
        \label{fig_pandax2_DMe_limits}
    \end{figure}

In \fig{fig_pandax2_DMe_limits}, We show the constraints on DM-electron scattering cross section with $f_l=0.1$ and $s=0.8$ using the constant charge yield model~\cite{Cheng:2021fqb,Aprile:2007qd} and the NEST2 charge yield model~\cite{Cheng:2021fqb,Szydagis:2020isq, NEST2calculator}.
We find that taking $s\simeq 0.8$ and $f_l\simeq 0.1$ makes our results most consistent with that of PandaX-II~\cite{Cheng:2021fqb} for both the constant and the NEST2 charge yield model.
Therefore, we take $s=0.8$ and $f_l=0.1$ as the nominal parameters for the analysis of the Migdal effect in this work.

\section{Details on data analysis procedures of XENON10}\label{appendix_XENON10}
In this appendix, we briefly review the data analysis procedures of XENON10 for electron recoils with MC simulation~\cite{Essig:2019xkx, Essig:2012yx, Essig:2017kqs} and cross-check our results with that of~\REF{Essig:2019xkx, Essig:2017kqs}.
A primary ionized electron can generate quanta $n_1$ from its kinetic energy and additional quanta $n_2$ from the de-excitation of the target atom in liquid xenon.
Part of these quanta can finally become the electron quanta with probability $f_e=\lra{1-r}/\lra{1+\lre{N_{\text{ex}}/N_{\text{ion}}}}$, where $r$ is the mean recombination probability and almost vanishes at low kinetic energies for electron recoils~\cite{Sorensen:2011bd, Essig:2012yx, Essig:2017kqs},
and $\lre{N_{\text{ex}}/N_{\text{ion}}}$ is the mean ratio of the number of photon quanta over that of electron quanta~\cite{Essig:2012yx, Essig:2017kqs}.
Moreover, the primary ionized electron itself can contribute as single electron quanta $n_e'$ with probability $1-r$.
The electron quanta can be extracted into xenon gas by the extraction electric field and induce scintillation which is known as the S2 signals.

For DM-electron scatterings,
we take $n_1=\text{floor}(T_e/W)$ and $n_2$ as the minimal additional quanta contributed by the de-excitation of each shell of xenon atoms~\cite{Essig:2012yx, Essig:2017kqs},
where $W=13.8~\text{eV}$~\cite{Essig:2012yx, Essig:2017kqs,Essig:2019xkx} is the average energy required to create a single quantum in liquid xenon,
and the function $\text{floor}(x)$ returns the largest integer less than or equal to $x$.
For the Migdal effect,
we take $n_1+n_2=\text{floor}\lra{E_\text{em}/W}$~\cite{Essig:2019xkx}.
We take $r=0$, $\lre{N_{\text{ex}}/N_{\text{ion}}}=0.2$, and $f_e=0.83$ from~\cite{Sorensen:2011bd, Essig:2012yx, Essig:2017kqs}.
The number of electron quanta is given by $n_e=n_e'+n_e''$, where $n_e''$ is generated by a binomial distribution
$
        \text{binom}\lra{n_1+n_2,\,f_e}.
    $
The number of the extracted electron quanta $N_e^{\prime\prime}$ is generated by another binomial distribution
$
        \text{binom}\lra{n_e,\,\epsilon_{\text{ex}t}}
    $,
where $\epsilon_{\text{ex}t}$ is the extraction efficiency taken as $100\,\%$ for XENON10~\cite{Essig:2019xkx}.
The signals generated in xenon gas ${\text{S2}_\text{raw}}$ are generated by a normal distribution
$
\text{norm}\lra{GN_e^{\prime\prime},\,\Delta G\sqrt{N_e^{\prime\prime}}}
    $,
where $G=27$ and $\Delta G=6.2$ are the gas gain for a single electron and its variation~\cite{Essig:2017kqs, Essig:2019xkx}, respectively.
The $\text{S2}_\text{raw}$ signals generated in xenon gas can be received by the detector as the $\text{S2}$ signals with a probability which is taken as the product of the efficiency extracted from the \texttt{QEdark} code~\cite{QEdark} and the acceptance $0.92$ taken from~\cite{Essig:2017kqs}.

With the data analysis procedures above,
we cross-check our constraints on DM-electron and DM-nucleon scattering cross section with~\REF{Essig:2017kqs} and~\REF{Essig:2019xkx}, respectively.
The constraints are derived with the Binned Poisson method~\cite{Savage:2008er, Green:2001xy} at $90\,\%$ C.L. using the S2-only data of XENON10 with 7 bins ranging from $14$ to $203$~PE~\cite{Essig:2017kqs, QEdark} with exposure of 15 $\text{kg} \cdot \text{day}$.
We neglect the background rates due to the lack of complete estimation of the background available from XENON10.

\begin{figure}[t]
    \centering
        \includegraphics[width=\figsizeTwo]{\plotsdir/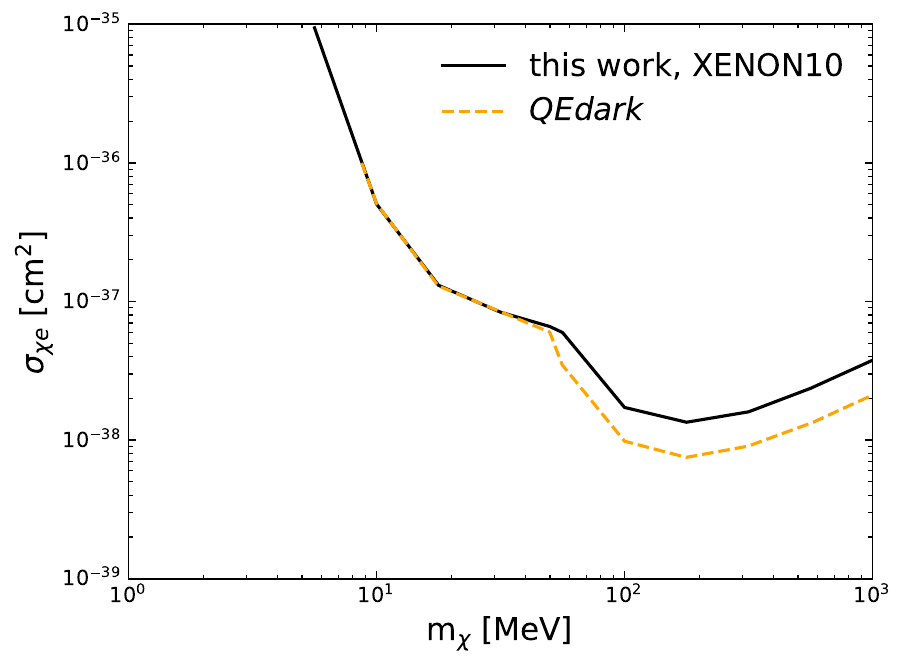}
        \includegraphics[width=\figsizeTwo]{\plotsdir/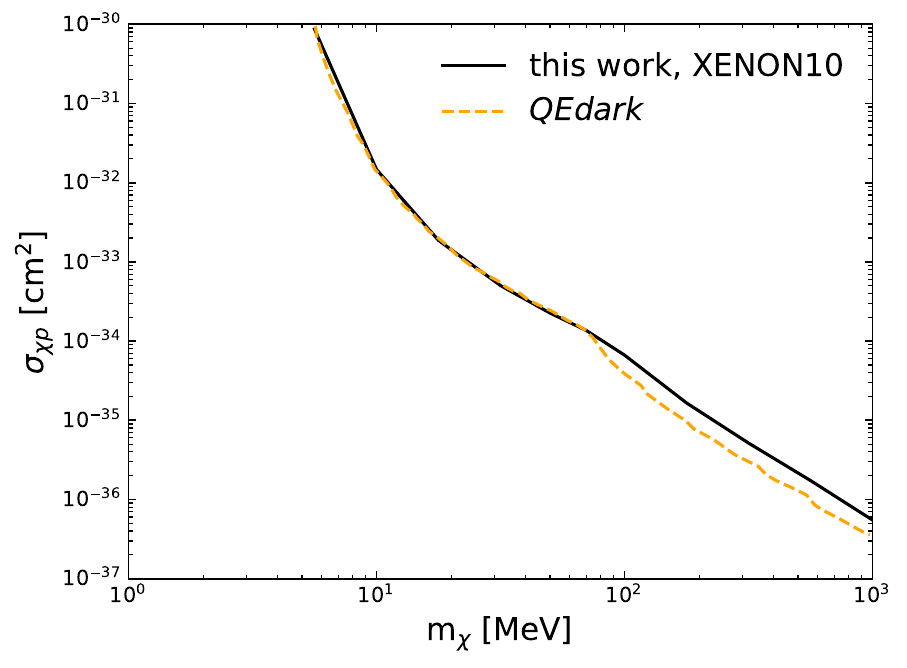}
    \caption{
        (Left)
Constraints on DM-electron scattering cross section~(black solid) at $90\,\%$ C.L. with the Binned Poisson method using the S2-only data of XENON10.
We show the result from~\cite{Essig:2017kqs}~(orange dashed) for comparison.
(Right)
Constraints on SI DM-nucleon scattering cross section~(black solid) at $90\,\%$ C.L. with the Binned Poisson method from the Migdal effect using the S2-only data of XENON10.
We also show the result from~\cite{Essig:2019xkx}~(orange dashed) for comparison.
        }
    \label{fig_Xenon10_limits}
\end{figure}
On the left panel of \fig{fig_Xenon10_limits},
we show the constraints on the DM-electron scattering cross section using the S2-only data of XENON10.
We take the DM local density and the speed of the Earth in the galactic rest frame as $0.4~\text{GeV}/\text{cm}^3$ and $227~\text{km/s}$, respectively, as in the \texttt{QEdark} code~\cite{QEdark}.
We note that the constraint set on DM-electron scattering cross section due to the data of XENON10 is calculated using individual bins as described in \REF{Essig:2017kqs}.
Therefore, our result is consistent with that of \REF{Essig:2017kqs} for $m_\chi \lesssim 50~\text{MeV}$ but a little weaker than that of \REF{Essig:2017kqs} for $m_\chi \gtrsim 50~\text{MeV}$ due to the different statistical inference methods.

On the right panel of \fig{fig_Xenon10_limits}, we show the constraints on the spin-independent~(SI) DM-nucleon scattering cross section from the Migdal effect using the S2-only data of XENON10. 
We take the DM local density of $0.3~\text{GeV}/\text{cm}^3$ as in ~\REF{Essig:2019xkx} and fix the speed of the Earth in galactic rest frame at $v_\oplus=232~\text{km/s}$~\cite{XENON:2018voc}.
Our result is consistent with that of \REF{Essig:2019xkx} for $m_\chi \lesssim 70~\text{MeV}$ but a little weaker than that of \REF{Essig:2019xkx} for $m_\chi \gtrsim 70~\text{MeV}$ for the same reason of DM-electron scattering case.

\section{Details on data analysis procedures of XENON1T}\label{appendix_XENON1T}
In this appendix, we briefly review the data analysis procedures for electron recoils of XENON1T based on the code~\cite{xenon1t:s2only_data_release},
where XENON1T collaboration has provided a response matrix including efficiency taken into account that transforms the spectra of the deposited electronic energy $E_{\text{em}}$ into S2 signals by the following integration
\begin{align}
        \frac{dR}{d{\text{S2}}}=\int_{E_\text{cut}}^\infty P({\text{S2}},~E_\text{em})\frac{dR}{dE_\text{em}}dE_\text{em},
    \end{align}
where $P({\text{S2}},~E_\text{em})$ is the response matrix provided in the code~\cite{xenon1t:s2only_data_release}
and $E_\text{cut}$ is a lower cutoff on $E_\text{em}$.
In \REF{Aprile:2019xxb}, the XENON1T collaboration has considered the cases with or without a lower cutoff on $E_\text{em}$ at $0.186~{\text{keV}}$.
We consider both of these two cases in this appendix.
The region of interest is from $150$ to $3000$~PE, which is divided into 12 bins evenly in the geometric space~\cite{Aprile:2019xxb}.
We only count the electron recoil background and coherent neutrino-nucleus scattering background~($\text{CE}\nu\text{NS}$) from the code~\cite{xenon1t:s2only_data_release}
and ignore the cathode background which is estimated from a data-driven method~\cite{xenon1t:s2only_data_release,Aprile:2019xxb} to get a conservative constraint.
The constraints are derived with the Binned Poisson method~\cite{Savage:2008er, Green:2001xy} at $90\,\%$ C.L. with effective exposure of 22~$\text{ton} \cdot \text{day}$.

\begin{figure}[t]
    \centering
        \includegraphics[width=\figsizeOne]{\plotsdir/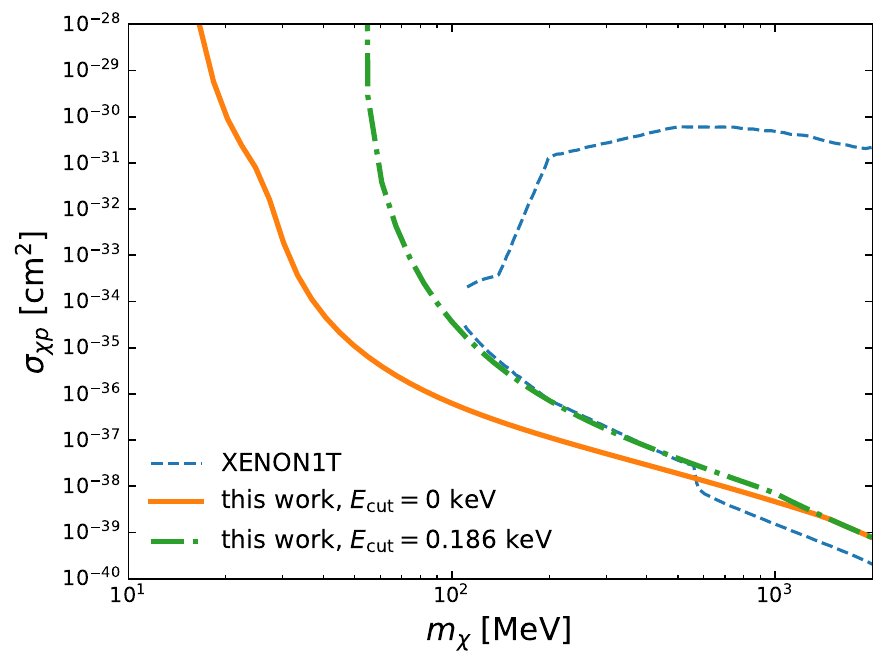}
    \caption{Constraints on SI DM-nucleon scattering cross section at $90\,\%$ C.L. from the Migdal effect using the S2-only data of XENON1T~\cite{xenon1t:s2only_data_release} with $E_\text{cut}=0~{\text{keV}}$~(orange solid) and $E_\text{cut}=0.186~{\text{keV}}$~(green dash-dotted).
The result of XENON1T collaboration~\cite{Aprile:2019jmx}~(blue dashed) is also shown for comparison.}\label{fig_reproduce_XENON1T}
\end{figure}

In \fig{fig_reproduce_XENON1T},
we show the constraints on SI DM-nucleon scattering cross section from the Migdal effect using S2-only data of XENON1T without the Earth shielding effect.
With $E_\text{cut}=0.186~{\text{keV}}$, the constraints with DM mass from 100~MeV to 500~MeV are consistent with the result of XENON1T~\cite{Aprile:2019jmx}.
The discrepancy above 500~MeV may be caused by the differences in the statistical inference methods.
With $E_\text{cut}=0$, the S2-only data of XENON1T can constrain SI DM-nucleon scattering cross section with DM mass down to $\sim 20~\text{MeV}$ through the Migdal effect.

\bibliographystyle{arxivref}
\bibliography{bib}

\providecommand{\href}[2]{#2}\begingroup\raggedright\begin{thebibliography}{10}

\bibitem{Collar:1992qc}
J.~I. Collar and F.~T. Avignone, ``{Diurnal modulation effects in cold dark
  matter experiments},''
  \href{http://dx.doi.org/10.1016/0370-2693(92)90873-3}{{\em Phys. Lett. B}
  {\bfseries  275} (1992) 181--185}.

\bibitem{Collar:1993ss}
J.~I. Collar and F.~T. Avignone, III, ``{The Effect of elastic scattering in
  the Earth on cold dark matter experiments},''
  \href{http://dx.doi.org/10.1103/PhysRevD.47.5238}{{\em Phys. Rev. D}
  {\bfseries  47} (1993) 5238--5246}.

\bibitem{LUX:2018xvj}
{\bfseries  LUX} Collaboration, D.~S. Akerib {\em et~al.}, ``{Search for annual
  and diurnal rate modulations in the LUX experiment},''
  \href{http://dx.doi.org/10.1103/PhysRevD.98.062005}{{\em Phys. Rev. D}
  {\bfseries  98} no.~6, (2018) 062005},
  \href{http://arxiv.org/abs/1807.07113}{{\ttfamily arXiv:1807.07113
  [astro-ph.CO]}}.

\bibitem{DAMA-LIBRA:2014lld}
{\bfseries  DAMA-LIBRA} Collaboration, R.~Bernabei {\em et~al.}, ``{Model
  independent result on possible diurnal effect in DAMA/LIBRA-phase1},''
  \href{http://dx.doi.org/10.1140/epjc/s10052-014-2827-1}{{\em Eur. Phys. J. C}
  {\bfseries  74} no.~3, (2014) 2827},
  \href{http://arxiv.org/abs/1403.4733}{{\ttfamily arXiv:1403.4733
  [astro-ph.GA]}}.

\bibitem{Foot:2014osa}
R.~Foot and S.~Vagnozzi, ``{Diurnal modulation signal from dissipative hidden
  sector dark matter},''
  \href{http://dx.doi.org/10.1016/j.physletb.2015.06.063}{{\em Phys. Lett. B}
  {\bfseries  748} (2015) 61--66},
  \href{http://arxiv.org/abs/1412.0762}{{\ttfamily arXiv:1412.0762 [hep-ph]}}.

\bibitem{Chen:2021ifo}
Y.~Chen, B.~Fornal, P.~Sandick, J.~Shu, X.~Xue, Y.~Zhao, and J.~Zong, ``{Earth
  shielding and daily modulation from electrophilic boosted dark particles},''
  \href{http://dx.doi.org/10.1103/PhysRevD.107.033006}{{\em Phys. Rev. D}
  {\bfseries  107} no.~3, (2023) 033006},
  \href{http://arxiv.org/abs/2110.09685}{{\ttfamily arXiv:2110.09685
  [hep-ph]}}.

\bibitem{Bernabei:2015nia}
R.~Bernabei {\em et~al.}, ``{Investigating Earth shadowing effect with
  DAMA/LIBRA-phase1},''
  \href{http://dx.doi.org/10.1140/epjc/s10052-015-3473-y}{{\em Eur. Phys. J. C}
  {\bfseries  75} no.~5, (2015) 239},
  \href{http://arxiv.org/abs/1505.05336}{{\ttfamily arXiv:1505.05336
  [hep-ph]}}.

\bibitem{DarkSide-50:2022qzh}
{\bfseries  DarkSide-50} Collaboration, P.~Agnes {\em et~al.}, ``{Search for
  low-mass dark matter WIMPs with 12~ton-day exposure of DarkSide-50},''
  \href{http://dx.doi.org/10.1103/PhysRevD.107.063001}{{\em Phys. Rev. D}
  {\bfseries  107} no.~6, (2023) 063001},
  \href{http://arxiv.org/abs/2207.11966}{{\ttfamily arXiv:2207.11966
  [hep-ex]}}.

\bibitem{LZ:2022ufs}
{\bfseries  LZ} Collaboration, J.~Aalbers {\em et~al.}, ``{First Dark Matter
  Search Results from the LUX-ZEPLIN (LZ) Experiment},''
  \href{http://arxiv.org/abs/2207.03764}{{\ttfamily arXiv:2207.03764
  [hep-ex]}}.

\bibitem{PandaX-4T:2021bab}
{\bfseries  PandaX-4T} Collaboration, Y.~Meng {\em et~al.}, ``{Dark Matter
  Search Results from the PandaX-4T Commissioning Run},''
  \href{http://dx.doi.org/10.1103/PhysRevLett.127.261802}{{\em Phys. Rev.
  Lett.} {\bfseries  127} no.~26, (2021) 261802},
  \href{http://arxiv.org/abs/2107.13438}{{\ttfamily arXiv:2107.13438
  [hep-ex]}}.

\bibitem{XENON:2023sxq}
{\bfseries  XENON} Collaboration, E.~Aprile {\em et~al.}, ``{First Dark Matter
  Search with Nuclear Recoils from the XENONnT Experiment},''
  \href{http://arxiv.org/abs/2303.14729}{{\ttfamily arXiv:2303.14729
  [hep-ex]}}.

\bibitem{Rudnick:2003xyz}
R.~Rudnick and S.~Gao,
  \href{http://dx.doi.org/https://doi.org/10.1016/B0-08-043751-6/03016-4}{{\em
  Composition of the Continental Crust}}.
\newblock Pergamon, Oxford, 2003.

\bibitem{Barak:2020fql}
{\bfseries  SENSEI} Collaboration, L.~Barak {\em et~al.}, ``{SENSEI:
  Direct-Detection Results on sub-GeV Dark Matter from a New Skipper-CCD},''
  \href{http://dx.doi.org/10.1103/PhysRevLett.125.171802}{{\em Phys. Rev.
  Lett.} {\bfseries  125} no.~17, (2020) 171802},
  \href{http://arxiv.org/abs/2004.11378}{{\ttfamily arXiv:2004.11378
  [astro-ph.CO]}}.

\bibitem{An:2017ojc}
H.~An, M.~Pospelov, J.~Pradler, and A.~Ritz, ``{Directly Detecting MeV-scale
  Dark Matter via Solar Reflection},''
  \href{http://dx.doi.org/10.1103/PhysRevLett.120.141801}{{\em Phys. Rev.
  Lett.} {\bfseries  120} no.~14, (2018) 141801},
  \href{http://arxiv.org/abs/1708.03642}{{\ttfamily arXiv:1708.03642
  [hep-ph]}}. [Erratum: Phys.Rev.Lett. 121, 259903 (2018)].

\bibitem{Xia:2022tid}
C.~Xia, Y.-H. Xu, and Y.-F. Zhou, ``{Azimuthal asymmetry in cosmic-ray boosted
  dark matter flux},''
  \href{http://dx.doi.org/10.1103/PhysRevD.107.055012}{{\em Phys. Rev. D}
  {\bfseries  107} no.~5, (2023) 055012},
  \href{http://arxiv.org/abs/2206.11454}{{\ttfamily arXiv:2206.11454
  [hep-ph]}}.

\bibitem{Migdal}
A.~B. Migdal, ``{Ionization of atoms accompanying $\alpha$- and
  $\beta$-decay},'' {\em J. Phys. Acad. Sci. USSR} {\bfseries  4(1-6)} (1941)
  449--453.

\bibitem{Dolan:2017xbu}
M.~J. Dolan, F.~Kahlhoefer, and C.~McCabe, ``{Directly detecting sub-GeV dark
  matter with electrons from nuclear scattering},''
  \href{http://dx.doi.org/10.1103/PhysRevLett.121.101801}{{\em Phys. Rev.
  Lett.} {\bfseries  121} no.~10, (2018) 101801},
  \href{http://arxiv.org/abs/1711.09906}{{\ttfamily arXiv:1711.09906
  [hep-ph]}}.

\bibitem{Ibe:2017yqa}
M.~Ibe, W.~Nakano, Y.~Shoji, and K.~Suzuki, ``{Migdal Effect in Dark Matter
  Direct Detection Experiments},''
  \href{http://dx.doi.org/10.1007/JHEP03(2018)194}{{\em JHEP} {\bfseries  03}
  (2018) 194}, \href{http://arxiv.org/abs/1707.07258}{{\ttfamily
  arXiv:1707.07258 [hep-ph]}}.

\bibitem{Essig:2019xkx}
R.~Essig, J.~Pradler, M.~Sholapurkar, and T.-T. Yu, ``{Relation between the
  Migdal Effect and Dark Matter-Electron Scattering in Isolated Atoms and
  Semiconductors},''
  \href{http://dx.doi.org/10.1103/PhysRevLett.124.021801}{{\em Phys. Rev.
  Lett.} {\bfseries  124} no.~2, (2020) 021801},
  \href{http://arxiv.org/abs/1908.10881}{{\ttfamily arXiv:1908.10881
  [hep-ph]}}.

\bibitem{Bernabei}
R.~Bernabei {\em et~al.}, ``{On electromagnetic contributions in WIMP
  quests},'' \href{http://dx.doi.org/10.1142/S0217751X07037093}{{\em Int. J.
  Mod. Phys. A} {\bfseries  22} (2007) 3155--3168},
  \href{http://arxiv.org/abs/0706.1421}{{\ttfamily arXiv:0706.1421
  [astro-ph]}}.

\bibitem{Aprile:2019xxb}
{\bfseries  XENON} Collaboration, E.~Aprile {\em et~al.}, ``{Light Dark Matter
  Search with Ionization Signals in XENON1T},''
  \href{http://dx.doi.org/10.1103/PhysRevLett.123.251801}{{\em Phys. Rev.
  Lett.} {\bfseries  123} no.~25, (2019) 251801},
  \href{http://arxiv.org/abs/1907.11485}{{\ttfamily arXiv:1907.11485
  [hep-ex]}}.

\bibitem{Li:2022acp}
J.~Li, L.~Su, L.~Wu, and B.~Zhu, ``{Spin-dependent sub-GeV inelastic dark
  matter-electron scattering and Migdal effect. Part I. Velocity independent
  operator},'' \href{http://dx.doi.org/10.1088/1475-7516/2023/04/020}{{\em
  JCAP} {\bfseries  04} (2023) 020},
  \href{http://arxiv.org/abs/2210.15474}{{\ttfamily arXiv:2210.15474
  [hep-ph]}}.

\bibitem{Flambaum:2020xxo}
V.~V. Flambaum, L.~Su, L.~Wu, and B.~Zhu, ``{New strong bounds on sub-GeV dark
  matter from boosted and Migdal effects},''
  \href{http://dx.doi.org/10.1007/s11433-022-2090-7}{{\em Sci. China Phys.
  Mech. Astron.} {\bfseries  66} no.~7, (2023) 271011},
  \href{http://arxiv.org/abs/2012.09751}{{\ttfamily arXiv:2012.09751
  [hep-ph]}}.

\bibitem{Cox:2022ekg}
P.~Cox, M.~J. Dolan, C.~McCabe, and H.~M. Quiney, ``{Precise predictions and
  new insights for atomic ionization from the Migdal effect},''
  \href{http://dx.doi.org/10.1103/PhysRevD.107.035032}{{\em Phys. Rev. D}
  {\bfseries  107} no.~3, (2023) 035032},
  \href{http://arxiv.org/abs/2208.12222}{{\ttfamily arXiv:2208.12222
  [hep-ph]}}.

\bibitem{Essig:2012yx}
R.~Essig, A.~Manalaysay, J.~Mardon, P.~Sorensen, and T.~Volansky, ``{First
  Direct Detection Limits on sub-GeV Dark Matter from XENON10},''
  \href{http://dx.doi.org/10.1103/PhysRevLett.109.021301}{{\em Phys. Rev.
  Lett.} {\bfseries  109} (2012) 021301},
  \href{http://arxiv.org/abs/1206.2644}{{\ttfamily arXiv:1206.2644
  [astro-ph.CO]}}.

\bibitem{DarkSide:2018ppu}
{\bfseries  DarkSide} Collaboration, P.~Agnes {\em et~al.}, ``{Constraints on
  Sub-GeV Dark-Matter\textendash{}Electron Scattering from the DarkSide-50
  Experiment},'' \href{http://dx.doi.org/10.1103/PhysRevLett.121.111303}{{\em
  Phys. Rev. Lett.} {\bfseries  121} no.~11, (2018) 111303},
  \href{http://arxiv.org/abs/1802.06998}{{\ttfamily arXiv:1802.06998
  [astro-ph.CO]}}.

\bibitem{Essig:2017kqs}
R.~Essig, T.~Volansky, and T.-T. Yu, ``{New Constraints and Prospects for
  sub-GeV Dark Matter Scattering off Electrons in Xenon},''
  \href{http://dx.doi.org/10.1103/PhysRevD.96.043017}{{\em Phys. Rev. D}
  {\bfseries  96} no.~4, (2017) 043017},
  \href{http://arxiv.org/abs/1703.00910}{{\ttfamily arXiv:1703.00910
  [hep-ph]}}.

\bibitem{Bunge:1993jsz}
C.~F. Bunge, J.~A. Barrientos, and A.~V. Bunge, ``{Roothaan-Hartree-Fock
  Ground-State Atomic Wave Functions: Slater-Type Orbital Expansions and
  Expectation Values for Z = 2-54},''
  \href{http://dx.doi.org/10.1006/adnd.1993.1003}{{\em Atom. Data Nucl. Data
  Tabl.} {\bfseries  53} (1993) 113--162}.

\bibitem{FAC}
M.~F. Gu, ``{The flexible atomic code},''
  \href{http://dx.doi.org/10.1139/p07-197}{{\em Canadian Journal of Physics}
  {\bfseries  86} no.~5, (2008) 675--689}.

\bibitem{Baxter:2019pnz}
D.~Baxter, Y.~Kahn, and G.~Krnjaic, ``{Electron Ionization via Dark
  Matter-Electron Scattering and the Migdal Effect},''
  \href{http://dx.doi.org/10.1103/PhysRevD.101.076014}{{\em Phys. Rev. D}
  {\bfseries  101} no.~7, (2020) 076014},
  \href{http://arxiv.org/abs/1908.00012}{{\ttfamily arXiv:1908.00012
  [hep-ph]}}.

\bibitem{Lewin:1995rx}
J.~D. Lewin and P.~F. Smith, ``{Review of mathematics, numerical factors, and
  corrections for dark matter experiments based on elastic nuclear recoil},''
  \href{http://dx.doi.org/10.1016/S0927-6505(96)00047-3}{{\em Astropart. Phys.}
  {\bfseries  6} (1996) 87--112}.

\bibitem{Helm:1956zz}
R.~H. Helm, ``{Inelastic and Elastic Scattering of 187-Mev Electrons from
  Selected Even-Even Nuclei},''
  \href{http://dx.doi.org/10.1103/PhysRev.104.1466}{{\em Phys. Rev.} {\bfseries
   104} (1956) 1466--1475}.

\bibitem{Engel:1991wq}
J.~Engel, ``{Nuclear form-factors for the scattering of weakly interacting
  massive particles},''
  \href{http://dx.doi.org/10.1016/0370-2693(91)90712-Y}{{\em Phys. Lett. B}
  {\bfseries  264} (1991) 114--119}.

\bibitem{jelle_aalbers_2022_7041453}
J.~Aalbers, B.~Pelssers, J.~R. Angevaare, and K.~D. Morå,
  ``Jelleaalbers/wimprates: v0.4.1,'' Sept., 2022.
\newblock \url{https://doi.org/10.5281/zenodo.7041453}.

\bibitem{Aprile:2019jmx}
{\bfseries  XENON} Collaboration, E.~Aprile {\em et~al.}, ``{Search for Light
  Dark Matter Interactions Enhanced by the Migdal Effect or Bremsstrahlung in
  XENON1T},'' \href{http://dx.doi.org/10.1103/PhysRevLett.123.241803}{{\em
  Phys. Rev. Lett.} {\bfseries  123} no.~24, (2019) 241803},
  \href{http://arxiv.org/abs/1907.12771}{{\ttfamily arXiv:1907.12771
  [hep-ex]}}.

\bibitem{Xu:2023wev}
J.~Xu {\em et~al.}, ``{Search for the Migdal effect in liquid xenon with
  keV-level nuclear recoils},''
  \href{http://arxiv.org/abs/2307.12952}{{\ttfamily arXiv:2307.12952
  [hep-ex]}}.

\bibitem{Dziewonski:1981xy}
A.~M. Dziewonski and D.~L. Anderson, ``{Preliminary reference earth model},''
  \href{http://dx.doi.org/10.1016/0031-9201(81)90046-7}{{\em Phys. Earth
  Planet. Interiors} {\bfseries  25} (1981) 297--356}.

\bibitem{Emken:2017qmp}
T.~Emken and C.~Kouvaris, ``{DaMaSCUS: The Impact of Underground Scatterings on
  Direct Detection of Light Dark Matter},''
  \href{http://dx.doi.org/10.1088/1475-7516/2017/10/031}{{\em JCAP} {\bfseries
  10} (2017) 031}, \href{http://arxiv.org/abs/1706.02249}{{\ttfamily
  arXiv:1706.02249 [hep-ph]}}.

\bibitem{ParticleDataGroup:2022pth}
{\bfseries  Particle Data Group} Collaboration, R.~L. Workman {\em et~al.},
  ``{Review of Particle Physics},''
  \href{http://dx.doi.org/10.1093/ptep/ptac097}{{\em PTEP} {\bfseries  2022}
  (2022) 083C01}.

\bibitem{Xia:2021vbz}
C.~Xia, Y.-H. Xu, and Y.-F. Zhou, ``{Production and attenuation of cosmic-ray
  boosted dark matter},''
  \href{http://dx.doi.org/10.1088/1475-7516/2022/02/028}{{\em JCAP} {\bfseries
  02} no.~02, (2022) 028}, \href{http://arxiv.org/abs/2111.05559}{{\ttfamily
  arXiv:2111.05559 [hep-ph]}}.

\bibitem{Super-Kamiokande:2022ncz}
{\bfseries  Super-Kamiokande} Collaboration, K.~Abe {\em et~al.}, ``{Search for
  Cosmic-Ray Boosted Sub-GeV Dark Matter Using Recoil Protons at
  Super-Kamiokande},''
  \href{http://dx.doi.org/10.1103/PhysRevLett.130.031802}{{\em Phys. Rev.
  Lett.} {\bfseries  130} no.~3, (2023) 031802},
  \href{http://arxiv.org/abs/2209.14968}{{\ttfamily arXiv:2209.14968
  [hep-ex]}}.

\bibitem{Kouvaris:2014lpa}
C.~Kouvaris and I.~M. Shoemaker, ``{Daily modulation as a smoking gun of dark
  matter with significant stopping rate},''
  \href{http://dx.doi.org/10.1103/PhysRevD.90.095011}{{\em Phys. Rev. D}
  {\bfseries  90} (2014) 095011},
  \href{http://arxiv.org/abs/1405.1729}{{\ttfamily arXiv:1405.1729 [hep-ph]}}.

\bibitem{Starkman:1990nj}
G.~D. Starkman, A.~Gould, R.~Esmailzadeh, and S.~Dimopoulos, ``{Opening the
  Window on Strongly Interacting Dark Matter},''
  \href{http://dx.doi.org/10.1103/PhysRevD.41.3594}{{\em Phys. Rev. D}
  {\bfseries  41} (1990) 3594}.

\bibitem{Kavanagh:2017cru}
B.~J. Kavanagh, ``{Earth scattering of superheavy dark matter: Updated
  constraints from detectors old and new},''
  \href{http://dx.doi.org/10.1103/PhysRevD.97.123013}{{\em Phys. Rev. D}
  {\bfseries  97} no.~12, (2018) 123013},
  \href{http://arxiv.org/abs/1712.04901}{{\ttfamily arXiv:1712.04901
  [hep-ph]}}.

\bibitem{Bringmann:2018cvk}
T.~Bringmann and M.~Pospelov, ``{Novel direct detection constraints on light
  dark matter},'' \href{http://dx.doi.org/10.1103/PhysRevLett.122.171801}{{\em
  Phys. Rev. Lett.} {\bfseries  122} no.~17, (2019) 171801},
  \href{http://arxiv.org/abs/1810.10543}{{\ttfamily arXiv:1810.10543
  [hep-ph]}}.

\bibitem{Xia:2020wcp}
C.~Xia, Y.-H. Xu, and Y.-F. Zhou, ``{Constraining light dark matter upscattered
  by ultrahigh-energy cosmic rays},''
  \href{http://dx.doi.org/10.1016/j.nuclphysb.2021.115470}{{\em Nucl. Phys. B}
  {\bfseries  969} (2021) 115470},
  \href{http://arxiv.org/abs/2009.00353}{{\ttfamily arXiv:2009.00353
  [hep-ph]}}.

\bibitem{Ge:2020yuf}
S.-F. Ge, J.~Liu, Q.~Yuan, and N.~Zhou, ``{Diurnal Effect of Sub-GeV Dark
  Matter Boosted by Cosmic Rays},''
  \href{http://dx.doi.org/10.1103/PhysRevLett.126.091804}{{\em Phys. Rev.
  Lett.} {\bfseries  126} no.~9, (2021) 091804},
  \href{http://arxiv.org/abs/2005.09480}{{\ttfamily arXiv:2005.09480
  [hep-ph]}}.

\bibitem{Cappiello:2019qsw}
C.~V. Cappiello and J.~F. Beacom, ``{Strong New Limits on Light Dark Matter
  from Neutrino Experiments},''
  \href{http://dx.doi.org/10.1103/PhysRevD.104.069901}{{\em Phys. Rev. D}
  {\bfseries  100} no.~10, (2019) 103011},
  \href{http://arxiv.org/abs/1906.11283}{{\ttfamily arXiv:1906.11283
  [hep-ph]}}. [Erratum: Phys.Rev.D 104, 069901 (2021)].

\bibitem{Emken:2018run}
T.~Emken and C.~Kouvaris, ``{How blind are underground and surface detectors to
  strongly interacting Dark Matter?},''
  \href{http://dx.doi.org/10.1103/PhysRevD.97.115047}{{\em Phys. Rev. D}
  {\bfseries  97} no.~11, (2018) 115047},
  \href{http://arxiv.org/abs/1802.04764}{{\ttfamily arXiv:1802.04764
  [hep-ph]}}.

\bibitem{CDEX:2021cll}
{\bfseries  CDEX} Collaboration, Z.~Z. Liu {\em et~al.}, ``{Studies of the
  Earth shielding effect to direct dark matter searches at the China Jinping
  Underground Laboratory},''
  \href{http://dx.doi.org/10.1103/PhysRevD.105.052005}{{\em Phys. Rev. D}
  {\bfseries  105} no.~5, (2022) 052005},
  \href{http://arxiv.org/abs/2111.11243}{{\ttfamily arXiv:2111.11243
  [hep-ex]}}.

\bibitem{PROSPECT:2021awi}
{\bfseries  PROSPECT, (PROSPECT Collaboration)*} Collaboration, M.~Andriamirado
  {\em et~al.}, ``{Limits on sub-GeV dark matter from the PROSPECT reactor
  antineutrino experiment},''
  \href{http://dx.doi.org/10.1103/PhysRevD.104.012009}{{\em Phys. Rev. D}
  {\bfseries  104} no.~1, (2021) 012009},
  \href{http://arxiv.org/abs/2104.11219}{{\ttfamily arXiv:2104.11219
  [hep-ex]}}.

\bibitem{Bringmann:2018lay}
T.~Bringmann, J.~Edsj\"o, P.~Gondolo, P.~Ullio, and L.~Bergstr\"om, ``{DarkSUSY
  6 : An Advanced Tool to Compute Dark Matter Properties Numerically},''
  \href{http://dx.doi.org/10.1088/1475-7516/2018/07/033}{{\em JCAP} {\bfseries
  07} (2018) 033}, \href{http://arxiv.org/abs/1802.03399}{{\ttfamily
  arXiv:1802.03399 [hep-ph]}}.

\bibitem{Kavanagh:2016pyr}
B.~J. Kavanagh, R.~Catena, and C.~Kouvaris, ``{Signatures of Earth-scattering
  in the direct detection of Dark Matter},''
  \href{http://dx.doi.org/10.1088/1475-7516/2017/01/012}{{\em JCAP} {\bfseries
  01} (2017) 012}, \href{http://arxiv.org/abs/1611.05453}{{\ttfamily
  arXiv:1611.05453 [hep-ph]}}.

\bibitem{Emken:2021lgc}
T.~Emken, ``{Solar reflection of light dark matter with heavy mediators},''
  \href{http://dx.doi.org/10.1103/PhysRevD.105.063020}{{\em Phys. Rev. D}
  {\bfseries  105} no.~6, (2022) 063020},
  \href{http://arxiv.org/abs/2102.12483}{{\ttfamily arXiv:2102.12483
  [hep-ph]}}.

\bibitem{Bramante:2022pmn}
J.~Bramante, J.~Kumar, G.~Mohlabeng, N.~Raj, and N.~Song, ``{Light Dark Matter
  Accumulating in Terrestrial Planets: Nuclear Scattering},''
  \href{http://arxiv.org/abs/2210.01812}{{\ttfamily arXiv:2210.01812
  [hep-ph]}}.

\bibitem{Cappiello:2023hza}
C.~V. Cappiello, ``{Analytic Approach to Light Dark Matter Propagation},''
  \href{http://dx.doi.org/10.1103/PhysRevLett.130.221001}{{\em Phys. Rev.
  Lett.} {\bfseries  130} no.~22, (2023) 221001},
  \href{http://arxiv.org/abs/2301.07728}{{\ttfamily arXiv:2301.07728
  [hep-ph]}}.

\bibitem{McDonough:2003}
W.~McDonough, ``Compositional model for the earth's core,''
  \href{http://dx.doi.org/10.1016/B0-08-043751-6/02015-6}{{\em Treatise on
  Geochemistry} {\bfseries  2} (11, 2003) 547--568}.

\bibitem{DarkSide:2022dhx}
{\bfseries  DarkSide} Collaboration, P.~Agnes {\em et~al.}, ``{Search for
  Dark-Matter\textendash{}Nucleon Interactions via Migdal Effect with
  DarkSide-50},'' \href{http://dx.doi.org/10.1103/PhysRevLett.130.101001}{{\em
  Phys. Rev. Lett.} {\bfseries  130} no.~10, (2023) 101001},
  \href{http://arxiv.org/abs/2207.11967}{{\ttfamily arXiv:2207.11967
  [hep-ex]}}.

\bibitem{Cheng:2021fqb}
{\bfseries  PandaX-II} Collaboration, C.~Cheng {\em et~al.}, ``{Search for
  Light Dark Matter-Electron Scatterings in the PandaX-II Experiment},''
  \href{http://dx.doi.org/10.1103/PhysRevLett.126.211803}{{\em Phys. Rev.
  Lett.} {\bfseries  126} no.~21, (2021) 211803},
  \href{http://arxiv.org/abs/2101.07479}{{\ttfamily arXiv:2101.07479
  [hep-ex]}}.

\bibitem{PandaX:2022xqx}
{\bfseries  PandaX} Collaboration, S.~Li {\em et~al.}, ``{Search for Light Dark
  Matter with Ionization Signals in the PandaX-4T Experiment},''
  \href{http://dx.doi.org/10.1103/PhysRevLett.130.261001}{{\em Phys. Rev.
  Lett.} {\bfseries  130} no.~26, (2023) 261001},
  \href{http://arxiv.org/abs/2212.10067}{{\ttfamily arXiv:2212.10067
  [hep-ex]}}.

\bibitem{PandaX-II:2020oim}
{\bfseries  PandaX-II} Collaboration, Q.~Wang {\em et~al.}, ``{Results of dark
  matter search using the full PandaX-II exposure},''
  \href{http://dx.doi.org/10.1088/1674-1137/abb658}{{\em Chin. Phys. C}
  {\bfseries  44} no.~12, (2020) 125001},
  \href{http://arxiv.org/abs/2007.15469}{{\ttfamily arXiv:2007.15469
  [astro-ph.CO]}}.

\bibitem{PandaX-II:2021jmq}
{\bfseries  PandaX-II} Collaboration, B.~Yan {\em et~al.}, ``{Determination of
  responses of liquid xenon to low energy electron and nuclear recoils using a
  PandaX-II detector},'' \href{http://dx.doi.org/10.1088/1674-1137/abf6c2}{{\em
  Chin. Phys. C} {\bfseries  45} no.~7, (2021) 075001},
  \href{http://arxiv.org/abs/2102.09158}{{\ttfamily arXiv:2102.09158
  [physics.ins-det]}}.

\bibitem{Sorensen:2011bd}
P.~Sorensen and C.~E. Dahl, ``{Nuclear recoil energy scale in liquid xenon with
  application to the direct detection of dark matter},''
  \href{http://dx.doi.org/10.1103/PhysRevD.83.063501}{{\em Phys. Rev. D}
  {\bfseries  83} (2011) 063501},
  \href{http://arxiv.org/abs/1101.6080}{{\ttfamily arXiv:1101.6080
  [astro-ph.IM]}}.

\bibitem{Zhang:2022wzy}
D.~Zhang, \href{http://dx.doi.org/10.13016/xeo9-zz9r}{{\em {Searches on Weakly
  Interacting Massive Particles and sub-MeV Fermionic Dark Matter in PandaX-II
  and PandaX-4T}}}.
\newblock PhD thesis, Maryland U., 2022.

\bibitem{xenon1t:s2only_data_release}
X.~Collaboration, ``{XENON1T/s2only\_data\_release: XENON1T S2-only data
  release},''. \url{https://doi.org/10.5281/zenodo.4075018}.

\bibitem{Savage:2008er}
C.~Savage, G.~Gelmini, P.~Gondolo, and K.~Freese, ``{Compatibility of
  DAMA/LIBRA dark matter detection with other searches},''
  \href{http://dx.doi.org/10.1088/1475-7516/2009/04/010}{{\em JCAP} {\bfseries
  04} (2009) 010}, \href{http://arxiv.org/abs/0808.3607}{{\ttfamily
  arXiv:0808.3607 [astro-ph]}}.

\bibitem{Green:2001xy}
A.~M. Green, ``{Calculating exclusion limits for weakly interacting massive
  particle direct detection experiments without background subtraction},''
  \href{http://dx.doi.org/10.1103/PhysRevD.65.023520}{{\em Phys. Rev. D}
  {\bfseries  65} (2002) 023520},
  \href{http://arxiv.org/abs/astro-ph/0106555}{{\ttfamily
  arXiv:astro-ph/0106555}}.

\bibitem{QEdark}
``{QEdark}.''
\newblock \url{https://github.com/tientienyu/QEdark}.

\bibitem{XENON:2018voc}
{\bfseries  XENON} Collaboration, E.~Aprile {\em et~al.}, ``{Dark Matter Search
  Results from a One Ton-Year Exposure of XENON1T},''
  \href{http://dx.doi.org/10.1103/PhysRevLett.121.111302}{{\em Phys. Rev.
  Lett.} {\bfseries  121} no.~11, (2018) 111302},
  \href{http://arxiv.org/abs/1805.12562}{{\ttfamily arXiv:1805.12562
  [astro-ph.CO]}}.

\bibitem{McCabe:2013kea}
C.~McCabe, ``{The Earth's velocity for direct detection experiments},''
  \href{http://dx.doi.org/10.1088/1475-7516/2014/02/027}{{\em JCAP} {\bfseries
  02} (2014) 027}, \href{http://arxiv.org/abs/1312.1355}{{\ttfamily
  arXiv:1312.1355 [astro-ph.CO]}}.

\bibitem{timon_emken_2020_3726878}
T.~Emken, ``temken/damascus: Version 1.1,'' Mar., 2020.
\newblock \url{https://doi.org/10.5281/zenodo.3726878}.

\bibitem{Aprile:2007qd}
E.~Aprile, K.~L. Giboni, P.~Majewski, K.~Ni, and M.~Yamashita, ``{Observation
  of Anti-correlation between Scintillation and Ionization for MeV Gamma-Rays
  in Liquid Xenon},'' \href{http://dx.doi.org/10.1103/PhysRevB.76.014115}{{\em
  Phys. Rev. B} {\bfseries  76} (2007) 014115},
  \href{http://arxiv.org/abs/0704.1118}{{\ttfamily arXiv:0704.1118
  [astro-ph]}}.

\bibitem{Szydagis:2020isq}
M.~Szydagis, C.~Levy, G.~M. Blockinger, A.~Kamaha, N.~Parveen, and G.~R.~C.
  Rischbieter, ``{Investigating the XENON1T low-energy electronic recoil excess
  using NEST},'' \href{http://dx.doi.org/10.1103/PhysRevD.103.012002}{{\em
  Phys. Rev. D} {\bfseries  103} no.~1, (2021) 012002},
  \href{http://arxiv.org/abs/2007.00528}{{\ttfamily arXiv:2007.00528
  [hep-ex]}}.

\bibitem{NEST2calculator}
``{Noble Element Simulation Technique}.''
  \url{http://nest.physics.ucdavis.edu/download/calculator}.

\end{thebibliography}\endgroup
\end{document}